\documentclass[12pt,letterpaper]{article}

\usepackage[left=1in,right=1in,top=1in,bottom=1in]{geometry}
\usepackage{setspace} 

\usepackage{amsmath,amssymb,amsthm,mathtools,bm}

\usepackage{graphicx}
\usepackage{booktabs}
\usepackage{threeparttable}
\usepackage{longtable}
\usepackage{array}
\usepackage{multirow}
\usepackage{caption}
\usepackage{subcaption}
\usepackage{float}
\usepackage{placeins}
\usepackage{makecell}

\usepackage{enumitem}
\usepackage{xcolor}
\usepackage{appendix}
\usepackage{pdflscape}
\usepackage{csquotes}
\usepackage{bbm}

\usepackage{chngcntr}
\counterwithin{figure}{section}
\counterwithin{table}{section}

\usepackage[authoryear]{natbib}

\usepackage[colorlinks=true,linkcolor=blue,citecolor=blue,urlcolor=blue]{hyperref}
\usepackage[nameinlink,capitalize]{cleveref}

\theoremstyle{definition}

\title{Which Portfolios?\\[0.5em]
	\large The Construction Dependence of Factor Model Performance}

\author{Useong Shin\thanks{
		Sogang Business School, Sogang University (Seoul, Korea).\\
		ORCID: \href{https://orcid.org/0009-0003-0197-9003}{0009-0003-0197-9003}\\
		Email: \texttt{useong@sogang.ac.kr}
}}
\date{\today}

\begin{document}
	
	\maketitle
	\thispagestyle{empty}
	
	\begin{flushleft}
	\end{flushleft}
	
	
	\begin{abstract}
		Factor-model performance depends not only on the model but also on how test assets are constructed. We form characteristic-unsorted random portfolios from a broad CRSP universe and vary stock selection, initial weighting, holding, and rebalancing. Rankings shift materially: buy-and-hold favors FF5 and FF6, whereas daily constant-weighting favors FF3, the most stable model across designs. Although q5 attains the highest maximum Sharpe ratio in factor-spanning tests, it leaves comparatively large and construction-sensitive pricing errors on random portfolios. These results reflect construction-specific weighting of each model's pricing-error vector. Test-asset construction, including dynamic weight management, is therefore a design choice in model evaluation.
	\end{abstract}
	
	\pagenumbering{arabic}
	
	\newpage
	\section{Introduction}
	\label{sec:intro}
	
	The empirical performance of a factor model depends not only on the model, but also on the assets used to evaluate it. Cross-sectional asset-pricing studies commonly rely on characteristic-sorted portfolios because they reduce idiosyncratic noise and isolate economically interpretable return patterns. Yet candidate factors and test portfolios are often built from related sorting rules. Strong fit on one familiar test-asset family may therefore combine broad pricing ability with alignment to a particular construction scheme. Whether that fit generalizes to portfolios formed by a different rule is a separate empirical question.
	
	I study this question using characteristic-unsorted random portfolios formed from a broad CRSP investable universe. Stocks are sampled either in proportion to market equity or with equal probabilities, and selected stocks receive value, equal, square-root market-equity, or random initial weights. I also vary portfolio breadth. To separate static formation from dynamic management, the same selected stocks and initial weights are implemented as buy-and-hold (BH), daily constant-weight (CW), and intermediate-frequency rebalanced portfolios. Test-asset construction therefore includes not only stock selection and initial weighting, but also the post-formation path of portfolio weights.
	
	The analysis compares the CAPM, FF3, Carhart, FF5, FF6, and q5 in three complementary exercises. I first evaluate the models on conventional characteristic-sorted portfolios from the Kenneth R. French Data Library and global-q. I then conduct direct spanning tests among the traded factors. Finally, I measure pricing errors on random portfolios using signed and absolute alpha, root-mean-square and tail errors, the share of significant alphas, and time-series $R^2$.
	
	This sequence distinguishes two objectives that are often conflated. Under the maximum-Sharpe criterion of \citet{BS17}, the relevant comparison among traded factor models is whether one model spans the factors of another; pricing errors on a common test-asset set do not produce a test-asset-invariant ranking under that objective. The random-portfolio exercise instead evaluates return-unit pricing errors on portfolios generated by a prespecified construction rule. Its rankings describe conditional absolute fit and construction sensitivity, not universal model superiority or maximum-Sharpe dominance.
	
	The results differ sharply across these objectives and test assets. On conventional characteristic-sorted portfolios, models perform relatively well on assets associated with their own research lineage, a pattern I call home-field asymmetry. Direct spanning tests favor q5 as providing the broadest sample mean--variance opportunity set, although neither q5 nor FF6 fully spans the other in every specification.
	
	Random portfolios produce a different ranking. Under BH and low-frequency rebalancing, FF5, FF6, and Carhart generally perform strongly, with the exact leader depending on stock selection and initial weights. Under daily CW, FF3 has the smallest pricing errors in most designs and is the most stable model across holding and rebalancing rules. More frequent rebalancing also generates widespread positive intercepts for several models. q5 leaves comparatively large and construction-sensitive errors, especially under CW, and greater diversification does not eliminate them.
	
	Factor-ablation results show that these patterns are not mechanically driven by factor count. The marginal effects of q5's profitability and expected-growth factors vary across test-asset families and weight-management rules: a factor that reduces pricing errors in one design can increase them in another. The broader conclusion also survives NYSE-only samples, the exclusion of stocks below the NYSE twentieth-percentile market-equity breakpoint, and intermediate rebalancing frequencies, although individual error levels and rankings move.
	
	The paper makes three contributions. First, it develops a systematic random-portfolio design that samples alternative tradable directions from the CRSP return space without directly reproducing the characteristic sorts used to construct or validate the candidate models. Second, it separates factor spanning from portfolio-level absolute fit, showing empirically that a broader mean--variance opportunity set can coexist with larger pricing errors on a prespecified portfolio population. Third, it treats dynamic weight management as part of test-asset design: changing the holding or rebalancing rule changes the return process being priced and can alter both the magnitude and direction of estimated errors.
	
	The objective is not to identify a model that is superior in every sense. It is to document where different models fit or fail and how those conclusions depend on the assets used to evaluate them. Test-asset construction---both static and dynamic---is not a neutral preprocessing choice; it is part of the model-evaluation design.
	
	\section{Theoretical Background and Related Literature}
	\label{sec:lit}
	
	\subsection{Linear Factor Models and Pricing Errors}
	
	Under the stochastic discount factor (SDF) approach, a valid SDF must price all tradable assets in a common investment universe. A factor model can be viewed as a low-dimensional approximation to such an SDF \citep{Cochrane05}. Let $f_t$ denote a vector of tradable factor returns and $R_t^e$ the excess returns on $N$ test assets:
	\begin{equation}
		R_t^e
		=
		\boldsymbol{\alpha}
		+
		Bf_t
		+
		\boldsymbol{\varepsilon}_t,
		\label{eq:factor_time_series}
	\end{equation}
	where $\boldsymbol{\alpha}$ is the vector of pricing errors. An exact model requires
	$\boldsymbol{\alpha}=\boldsymbol{0}$ for every tradable asset. Empirical models are incomplete approximations and may therefore leave errors in different directions of the return space. Their measured performance depends not only on the factors, but also on which directions are selected as test assets and how the errors are aggregated.
	
	The joint-intercept test of \citet{GRS89}, for example, weights pricing errors by residual precision:
	\begin{equation}
		\mathrm{GRS}
		=
		\frac{T-N-K}{N}
		\frac{
			\widehat{\boldsymbol{\alpha}}^{\prime}
			\widehat{\Sigma}_{\varepsilon}^{-1}
			\widehat{\boldsymbol{\alpha}}
		}{
			1+
			\widehat{\boldsymbol{\mu}}_{f}^{\prime}
			\widehat{\Sigma}_{f}^{-1}
			\widehat{\boldsymbol{\mu}}_{f}
		},
		\label{eq:grs_stat}
	\end{equation}
	where $T$ is the sample length and $K$ the number of factors. The statistic is closely related to whether the remaining errors can be combined into an additional mean--variance investment opportunity, rather than to the return-unit error of a typical portfolio.
	
	The Hansen--Jagannathan distance similarly aggregates pricing-condition errors through the second moments of returns. If $g(b)=E[m_t(b)R_t]-p$, then
	\begin{equation}
		d_{HJ}^{2}
		=
		\min_b
		g(b)^{\prime}G^{-1}g(b),
		\label{eq:hj_distance}
	\end{equation}
	where $G$ is the second-moment matrix of the test assets \citep{HJ97}. For an exact model, the choice of test assets and weighting matrix is irrelevant. For an incomplete model, both determine which dimensions of misspecification receive the greatest weight.
	
	\subsection{Maximum Sharpe Ratios and Portfolio-Level Absolute Fit}
	
	\citet{BS17} show that, for tradable factor models, the maximum-Sharpe comparison is governed by factor spanning. If test assets $R_t$ are added to factors $f_t$,
	\begin{equation}
		\operatorname{SR}^{2}(f,R)
		-
		\operatorname{SR}^{2}(f)
		=
		\boldsymbol{\alpha}^{\prime}
		\Sigma_{\varepsilon}^{-1}
		\boldsymbol{\alpha}.
		\label{eq:bs_sharpe_identity}
	\end{equation}
	The right-hand side is the investment opportunity obtained by optimally combining the residual returns. Under this objective, the central question is whether one factor model spans another, not which model has the smallest unweighted test-asset alphas \citep{BS18,FF18}.
	
	Test-asset comparisons nevertheless answer a different and still useful question. Studies such as \citet{HMXZ21} evaluate models by the magnitudes and significance of pricing errors on a specified anomaly cross-section. Such exercises diagnose where a model fits or fails within that universe, but do not produce a test-asset-invariant ranking of maximum Sharpe ratios. Test assets also determine factor identification: \citet{GXZ25} show that factor strength depends on the exposures contained in the selected cross-section.
	
	Let $\mathcal{A}$ denote a prespecified set of test assets. Portfolio-level absolute fit can be summarized by
	\begin{align}
		\mathcal{L}_{1}(M;\mathcal{A})
		&=
		\frac{1}{N}
		\sum_{i=1}^{N}
		\left|\alpha_{i,M}\right|,
		\\
		\mathcal{L}_{2}(M;\mathcal{A})
		&=
		\left(
		\frac{1}{N}
		\sum_{i=1}^{N}
		\alpha_{i,M}^{2}
		\right)^{1/2}.
		\label{eq:raw_alpha_losses}
	\end{align}
	Average signed alpha, tail errors, and the share of significant alphas provide complementary information on the direction, concentration, and statistical prevalence of misspecification. These measures neither replace the maximum Sharpe ratio nor contradict equation~\eqref{eq:bs_sharpe_identity}. They evaluate the return-unit pricing errors left on a particular portfolio population.
	
	This distinction limits the interpretation of model rankings. A model may span competing factors and offer a broader mean--variance opportunity set while leaving relatively large errors on a prespecified cross-section. Conversely, smaller average errors on that cross-section do not imply a higher maximum Sharpe ratio. The two findings correspond to different objective functions.
	
	\subsection{Test-Asset Endogeneity and Home-Field Bias}
	
	Characteristic-sorted portfolios reduce firm-level noise and isolate economically interpretable return patterns, but they may also align closely with the characteristics used to discover, construct, or validate candidate factors. Strong fit within such a sorting space need not generalize to portfolios formed by another rule.
	
	\citet{LM90} show that using empirically observed characteristics in both portfolio construction and testing can introduce data-snooping bias. \citet{LNS10} show that the strong common-factor structure of size- and value-sorted portfolios can make high cross-sectional fit less informative than it appears. \citet{KNS18} emphasize the dominant covariance directions implied by approximate no-arbitrage, while \citet{GXZ25} connect test-asset choice directly to weak-factor identification. Although these studies address different problems, they share the implication that factor-model evaluation cannot be separated from test-asset design.
	
	I use the term ``home-field bias'' for a symmetric form of this dependence: a model performs relatively well on test assets close to the characteristic space emphasized in its construction or prominent validation exercises. The term does not imply intentional overfitting or deny the economic content of the factors. Either the Fama--French or q-factor lineage may benefit when evaluated on portfolios that emphasize its own characteristic space.
	
	The relevant question is therefore not which data library is correct, but whether performance in a familiar sorting space extends to portfolios generated by an external rule. Characteristic-sorted and externally generated portfolios should be viewed as complementary diagnostics.
	
	\subsection{Asset-Generating Distributions and Random Portfolios}
	
	The random portfolios in this paper remain inside the CRSP return space: they are long-only combinations of actual stocks from the same investable universe. What changes is the asset-generation rule. Rather than sorting on candidate-model characteristics, I sample stock combinations and weights from prespecified distributions. Because a valid SDF must price both individual stocks and their tradable combinations, these portfolios reveal economically relevant directions that conventional sorts may not emphasize.
	
	For a constant-weight (CW) portfolio with fixed weights $w$, excess returns and pricing errors satisfy
	\begin{equation}
		R_{p,t}^{e,CW}
		=
		w^{\prime}R_t^e,
		\qquad
		\alpha_{p,M}^{CW}
		=
		w^{\prime}\boldsymbol{\alpha}_M.
		\label{eq:cw_alpha_linearity}
	\end{equation}
	If $\mathcal{D}$ is an ex ante distribution over portfolio weights, the integrated squared pricing error is
	\begin{equation}
		\mathcal{L}_{\mathcal{D}}^{CW}(M)
		=
		E_{w\sim\mathcal{D}}
		\left[
		\left(
		w^{\prime}\boldsymbol{\alpha}_M
		\right)^2
		\right]
		=
		\boldsymbol{\alpha}_M^{\prime}
		Q_{\mathcal{D}}
		\boldsymbol{\alpha}_M,
		\qquad
		Q_{\mathcal{D}}
		=
		E_{w\sim\mathcal{D}}[ww^{\prime}].
		\label{eq:distribution_loss_lit}
	\end{equation}
	Stock-selection probabilities, portfolio breadth, and weighting rules jointly determine $Q_{\mathcal{D}}$. Different asset-generating distributions therefore apply different weights to the same underlying pricing-error vector.
	
	The matrix $Q_{\mathcal{D}}$ is not a unique welfare criterion. It is an evaluation measure induced by a prespecified population of tradable portfolios. The paper consequently does not propose one optimal distribution; it varies selection probabilities, initial weights, portfolio breadth, and investable universes to determine whether a result is specific to one design or recurs across several.
	
	Selection and weighting affect different stages of construction. Uniform selection followed by value weighting resembles a value-weighted random subsample of the market. Market-equity-proportional selection followed by value weighting emphasizes large stocks twice, first through inclusion and then through capital allocation.
	
	Buy-and-hold (BH) and intermediate rebalancing require a dynamic extension. Let $\pi$ denote a weight-management rule:
	\begin{equation}
		R_{p,t}^{e,\pi}
		=
		\left(w_{p,t-1}^{\pi}\right)^{\prime}R_t^e,
		\label{eq:dynamic_portfolio_return}
	\end{equation}
	where $w_{p,t-1}^{\pi}$ depends on past returns and the rebalancing rule. In general,
	\[
	\alpha_{p,M}^{\pi}
	\neq
	w_{p,0}^{\prime}\boldsymbol{\alpha}_M.
	\]
	If $\Pi$ is a distribution over weight-management rules, the corresponding evaluation object is
	\begin{equation}
		\mathcal{L}_{\Pi}^{dyn}(M)
		=
		E_{\pi\sim\Pi}
		\left[
		\alpha_M\!\left(R^{e,\pi}\right)^2
		\right].
		\label{eq:dynamic_distribution_loss}
	\end{equation}
	CW is the limiting case in which the static argument applies exactly. Under BH and intermediate rebalancing, the weight path itself generates a distinct tradable return process. Static formation and dynamic management are therefore both components of test-asset construction.
	
	\subsection{Competing Factor Models and Scope}
	
	The empirical analysis compares the CAPM, the Fama--French three- and five-factor models \citep{FF92,FF93,FF15}, the Carhart model \citep{JT93,Carhart97}, FF6, and the investment-based q5 model \citep{HXZ15,HMXZ19,HXZ20,HMXZ21,HMXZ24}. Fama--French factors and portfolios are obtained from the Kenneth R. French Data Library, while q5 factors and anomaly portfolios come from global-q \citep{FrenchDataLibrary,globalqFactors}.
	
	The two research lineages emphasize different economic mechanisms and characteristic-sorting procedures. Their conventional test assets remain essential for determining whether each model explains the patterns it was designed to address. The unresolved question is whether such performance generalizes to assets formed under a different construction principle.
	
	The paper therefore conducts three complementary comparisons. First, it evaluates pricing errors on conventional characteristic-sorted portfolios to document home-field asymmetry. Second, it uses direct factor-spanning tests to compare relative mean--variance opportunity sets. Third, it examines the level, direction, and stability of pricing errors on random portfolios under alternative stock-selection, initial-weighting, breadth, holding, and rebalancing rules.
	
	The objective is neither to discard characteristic-sorted portfolios nor to declare a universally superior model from random-portfolio alphas. It is to distinguish factor spanning from portfolio-level absolute fit and to show how both static formation and dynamic weight management enter the evaluation of incomplete factor models.
	
	\section{Data and Methodology}
	\label{sec:data}
	
	\subsection{Construction of the CRSP Investable Universe}
	
	I construct the investable universe from daily CRSP stock data. The sample runs from January 3, 1967, through December 31, 2024, and begins with common stocks listed on the NYSE, AMEX, and NASDAQ. The screening procedure is designed to retain nearly all economically relevant market capitalization and trading activity while preventing extremely small and illiquid stocks from dominating the results.
	
	Investability is updated at each month-end using market equity and trading liquidity. The month-end market equity of stock $i$ is
	\begin{equation}
		ME_{i,t}
		=
		\left|PRC_{i,t}\right|
		\times SHROUT_{i,t}
		\times 1{,}000.
	\end{equation}
	To implement the size screen, I rank stocks in descending order of market equity and calculate the cumulative share of aggregate market equity. For stock $i$, this share is
	\begin{equation}
		CumME_{i,t}
		=
		\frac{
			\sum_{j:ME_{j,t}\geq ME_{i,t}} ME_{j,t}
		}{
			\sum_j ME_{j,t}
		}.
	\end{equation}
	
	Trading liquidity is measured by average daily dollar volume over the 63 trading days ending at month-end:
	\begin{align}
		DVOL_{i,d}
		&=
		\left|PRC_{i,d}\right|\times VOL_{i,d},
		\\
		ADV63_{i,t}
		&=
		\frac{1}{N_{i,t}}
		\sum_{d\in\mathcal{W}_t}DVOL_{i,d},
	\end{align}
	where $\mathcal{W}_t$ is the set of the most recent 63 trading days through month-end $t$, and $N_{i,t}$ is the number of days in that window for which dollar volume is observed. Because nominal market size and trading volume change substantially over the sample, I use the monthly cross-sectional percentile of $ADV63$ rather than an absolute dollar-volume threshold.
	
	The screen uses hysteresis, with different thresholds for entry and exit. A stock already in the universe is removed if $CumME_{i,t}>0.999$ or if its $ADV63$ falls at or below the 2.5th percentile of the monthly cross-section. A stock outside the universe may enter only if $CumME_{i,t}\leq0.995$ and its $ADV63$ reaches at least the fifth percentile. The stricter entry rule reduces repeated entry and exit near the screening thresholds.
	
	Historical NASDAQ volume data are incomplete in part of the sample. For a NASDAQ stock, I therefore suspend the liquidity screen until 63 trading days have elapsed after volume first becomes observable. The market-equity screen continues to apply during this grace period. All screening variables use information available by the relevant month-end, and the resulting universe becomes effective on the first trading day of the following month. This timing convention prevents future information from entering the selection process.
	
	Across the full sample, the final universe contains an average of 77.6\% of the common stocks in the initial sample while preserving approximately 99.7\% of aggregate market equity and 63-day average dollar volume. At the end of December 2024, the screen retains 2,426 of 3,804 common stocks. The retained stocks account for 99.78\% of market equity and 99.28\% of dollar volume. The procedure therefore removes the extreme illiquid tail while leaving the economic mass of the market largely unchanged.
	
	\subsection{Random-Portfolio Construction}
	
	Random portfolios are formed from the investable universe at the end of each June and held from the first trading day of July through the end of the following June. Stock selection and initial weights use only information available at formation. Subsequent returns and future delisting outcomes do not enter the formation decision. Each portfolio therefore begins with the same information set, but different combinations of stock selection, initial weighting, and weight management generate distinct tradable return processes.
	
	\subsubsection{Stock Selection}
	
	Let $N_t$ denote the number of investable stocks at formation date $t$, and let $\rho$ denote the stock-selection ratio. A portfolio contains approximately $\rho N_t$ stocks. The selection ratios are
	\begin{equation}
		\rho
		\in
		\left\{
		0.005,\,
		0.01,\,
		0.05,\,
		0.10,\,
		0.25,\,
		0.50,\,
		0.75,\,
		0.90,\,
		1.00
		\right\}.
	\end{equation}
	For each selection ratio and sampling rule, I generate 500 portfolios. When $\rho=1$, the selected stock set is unique, so only one stock draw is generated for each corresponding design.
	
	Within each formation year and portfolio identifier, I assign a common random ordering of stocks across selection ratios. The stock set associated with a lower selection ratio is therefore nested within the stock set associated with a higher ratio for the same portfolio identifier. This nesting reduces the extent to which changes across selection ratios are contaminated by differences between unrelated Monte Carlo draws.
	
	Stocks are sampled without replacement. Under market-equity-proportional sampling (MEP), let $\mathcal{R}_{p,t}^{(\ell)}$ denote the set of stocks remaining immediately before the $\ell$th draw. The conditional probability of selecting stock $i$ is
	\begin{equation}
		\Pr\!\left(
		i\text{ is selected on draw }\ell
		\mid \mathcal{R}_{p,t}^{(\ell)}
		\right)
		=
		\frac{ME_{i,t}}
		{\sum_{j\in\mathcal{R}_{p,t}^{(\ell)}}ME_{j,t}}.
	\end{equation}
	Under uniform sampling (UNIF), all remaining stocks receive the same conditional selection probability. MEP therefore favors the inclusion of large stocks, whereas UNIF gives every stock the same ex ante opportunity to enter the portfolio regardless of market equity.
	
	\subsubsection{Initial Weights}
	
	Selected stocks receive one of four initial weighting schemes: equal weighting (EW), value weighting (VW), square-root market-equity weighting (SQME), or random weighting (RAND). Let $\mathcal{S}_{p,t}$ be the stock set of portfolio $p$ at formation date $t$, and let $K_{p,t}$ be the number of selected stocks. The initial weights are
	\begin{align}
		w_{i,p,t}^{\mathrm{EW}}
		&=
		\frac{1}{K_{p,t}},
		\\
		w_{i,p,t}^{\mathrm{VW}}
		&=
		\frac{ME_{i,t}}
		{\sum_{j\in\mathcal{S}_{p,t}}ME_{j,t}},
		\\
		w_{i,p,t}^{\mathrm{SQME}}
		&=
		\frac{\sqrt{ME_{i,t}}}
		{\sum_{j\in\mathcal{S}_{p,t}}\sqrt{ME_{j,t}}},
		\\
		u_{i,p,t}
		&\sim
		\mathrm{Unif}(0,1),
		\qquad
		w_{i,p,t}^{\mathrm{RAND}}
		=
		\frac{u_{i,p,t}}
		{\sum_{j\in\mathcal{S}_{p,t}}u_{j,p,t}}.
	\end{align}
	All portfolios are long-only, and the initial weights sum to one. Combining the two sampling rules with the four weighting schemes produces eight baseline construction designs.
	
	The sampling rule and the weighting rule govern different stages of portfolio construction. MEP and UNIF determine which stocks enter the portfolio. EW, VW, SQME, and RAND determine how capital is allocated among the selected stocks. This separation makes it possible to distinguish large-stock exposure arising at the selection stage from large-stock exposure arising at the weighting stage.
	
	\subsubsection{Weight-Management Rules}
	
	I apply two weight-management rules to the same selected stocks and initial weights: constant weight (CW) and buy and hold (BH). Let $w_{i,p,t}^{0}$ denote the initial weight assigned at formation, and let $R_{i,d}^{*}$ denote the daily return of stock $i$, including the delisting treatment described below. The CW portfolio return is
	\begin{equation}
		R_{p,d}^{\mathrm{CW}}
		=
		\sum_{i\in\mathcal{S}_{p,t}}
		w_{i,p,t}^{0}R_{i,d}^{*}.
	\end{equation}
	Under CW, the portfolio is rebalanced after each trading day to restore the initial target weights. Conditional on the initial weights, its daily return is therefore a linear combination of stock returns with fixed coefficients.
	
	Under BH, realized returns endogenously change subsequent portfolio weights. Let $d_0$ denote the first trading day of the holding period. The cumulative value of stock $i$ immediately before trading day $d$ is
	\begin{equation}
		A_{i,p,d-1}
		=
		w_{i,p,t}^{0}
		\prod_{s=d_0}^{d-1}
		\left(1+R_{i,s}^{*}\right).
	\end{equation}
	The BH weight applied on day $d$ and the corresponding portfolio return are
	\begin{align}
		w_{i,p,d-1}^{\mathrm{BH}}
		&=
		\frac{A_{i,p,d-1}}
		{\sum_{j\in\mathcal{S}_{p,t}}A_{j,p,d-1}},
		\\
		R_{p,d}^{\mathrm{BH}}
		&=
		\sum_{i\in\mathcal{S}_{p,t}}
		w_{i,p,d-1}^{\mathrm{BH}}R_{i,d}^{*}.
	\end{align}
	Equivalently, define the BH portfolio value as
	\begin{equation}
		V_{p,d}^{\mathrm{BH}}
		=
		\sum_{i\in\mathcal{S}_{p,t}}
		w_{i,p,t}^{0}
		\prod_{s=d_0}^{d}
		\left(1+R_{i,s}^{*}\right).
	\end{equation}
	The daily return can then be written as
	\begin{equation}
		R_{p,d}^{\mathrm{BH}}
		=
		\frac{V_{p,d}^{\mathrm{BH}}}
		{V_{p,d-1}^{\mathrm{BH}}}-1.
	\end{equation}
	
	Conditional on past information, the BH return on a given day is a linear combination of contemporaneous stock returns. The coefficients in that combination, however, are determined by products of past realized returns. The mapping from the full paths of constituent returns to the portfolio return process is therefore path-dependent and nonlinear. Even with identical initial weights, the ordering of stock-level gains and losses changes subsequent exposures and can change the model's final pricing error. CW is the opposite limiting case: it removes weight drift by restoring the target portfolio every day.
	
	This distinction is not merely computational. Benchmark test portfolios from the Kenneth R. French Data Library and global--q also allow portfolio values and weights to evolve with realized constituent returns between scheduled formation or reconstitution dates \citep{FrenchDataLibrary,globalqFactors}. Real-world funds likewise tend to rebalance periodically rather than continuously. Performance under BH or low-frequency rebalancing is therefore not an auxiliary robustness requirement. It is relevant to whether a model describes the return processes generated by commonly used test portfolios and realistic portfolio management. I treat CW as a controlled benchmark that isolates the effect of fixed weights and BH as the main economic benchmark that incorporates path-dependent weight drift.
	
	CW and BH use the same stocks and the same initial weights. Their return processes differ only because of post-formation weight management. Comparing the two therefore isolates the dynamic dimension of test-asset construction.
	
	\subsubsection{Delistings and Reproducibility}
	
	Stocks that will subsequently delist are not removed at formation. A stock that is investable at the formation date remains in the portfolio through its actual delisting date, and I use a return measure that incorporates the CRSP delisting return. Temporary missing returns before delisting are set to zero. After delisting, the associated account is invested at the risk-free rate until the next annual formation date. This procedure incorporates delisting losses while avoiding the ex post selection of stocks that survive the full holding period.
	
	All random draws and random weights use a fixed seed.%
	\footnote{The random seed is 20260614 for all stochastic procedures.}
	The resulting portfolio returns are applied to every candidate factor model. Differences across models therefore reflect the pricing errors that the models leave on the same return processes rather than differences in the test assets supplied to each model.
	
	\subsection{Factor Data and Model-Specific Benchmark Returns}
	
	Each candidate model is paired with the market factor and risk-free rate supplied with that model's factor data. For the CAPM, the Fama--French models, and the Carhart model, I use $Mkt-RF$ and $RF$ from the Kenneth R. French Data Library. For q5, I use $R_{MKT}$ and $R_F$ from global--q. The excess return of portfolio $p$ under model $M$ is
	\begin{equation}
		R_{p,d}^{e,M}
		=
		R_{p,d}-RF_{d}^{M},
	\end{equation}
	and the market factor in the corresponding regression is drawn from the same data source as the model's risk-free rate.
	
	This convention avoids mechanically disadvantaging q5 by combining it with the Fama--French benchmark returns. In practice, however, the market and risk-free series from the two sources are extremely similar.%
	\footnote{Over the common 14,598 trading days from January 3, 1967, through December 31, 2024, the correlation between global--q's $R_{MKT}$ and the French Data Library's $Mkt-RF$ is 0.99997. Their mean daily absolute difference is 0.0036 percentage points, or approximately 0.36 bp. The mean daily absolute difference between the two risk-free rates is 0.0022 percentage points, or approximately 0.22 bp.}
	Using source-specific inputs is therefore a conservative choice that preserves symmetry across model families. The empirical differences reported below are too large to be explained by the source of the market factor or the risk-free rate.
	
	\section{Factor-Model Spanning Tests}
	\label{sec:span}
	
	Before examining random portfolios, I compare the mean--variance opportunity sets of the tradable factor models. Under the maximum-Sharpe criterion of \citet{BS17}, the relevant question is whether one model spans the factors of another, not which model leaves the smallest unadjusted alphas on a common test-asset set. The spanning tests therefore provide a complementary benchmark to the subsequent portfolio-level analysis \citep{HMXZ19}.
	
	\subsection{Test Design}
	
	FF6 contains the market, size, value, profitability, investment, and momentum factors, whereas q5 contains the market, size, investment, return-on-equity, and expected-growth factors. To avoid including two nearly identical market series in the same regression, the mutual spanning tests use a common market factor. The baseline uses the Fama--French market factor; using the global--q market factor yields the same conclusions.
	
	Define the model-specific factor blocks as
	\begin{align*}
		g_t^{q}
		&=
		(ME_t,IA_t,ROE_t,EG_t)^{\prime},
		\\
		g_t^{FF}
		&=
		(SMB_t,HML_t,RMW_t,CMA_t,MOM_t)^{\prime}.
	\end{align*}
	The mutual spanning regressions are
	\begin{align}
		g_t^{q}
		&=
		\boldsymbol{\alpha}_{q\mid FF6}
		+
		B_{q\mid FF6}f_t^{FF6}
		+
		\boldsymbol{u}_{q,t},
		\label{eq:span_ff6_q5}
		\\
		g_t^{FF}
		&=
		\boldsymbol{\alpha}_{FF\mid q5}
		+
		B_{FF\mid q5}f_t^{q5}
		+
		\boldsymbol{u}_{FF,t}.
		\label{eq:span_q5_ff6}
	\end{align}
	The null hypotheses are
	$H_0:\boldsymbol{\alpha}_{q\mid FF6}=\boldsymbol{0}$ and
	$H_0:\boldsymbol{\alpha}_{FF\mid q5}=\boldsymbol{0}$.
	I evaluate them using the GRS test and a Newey--West HAC Wald test, with lags of 21 trading days for daily data and six months for monthly data.
	
	The common sample contains 14,598 daily observations from January 3, 1967, through December 31, 2024, and 696 monthly observations obtained by compounding daily returns within calendar months. The excluded factor block's contribution to the opportunity set is summarized by
	\begin{equation}
		\Delta\operatorname{SR}^{2}
		=
		\widehat{\boldsymbol{\alpha}}^{\prime}
		\widehat{\Sigma}_{u}^{-1}
		\widehat{\boldsymbol{\alpha}},
		\label{eq:span_delta_sr2}
	\end{equation}
	where $\widehat{\Sigma}_{u}$ is the residual covariance matrix. Maximum Sharpe ratios are descriptive in-sample measures; formal spanning conclusions rely on the joint-intercept tests.
	
	\subsection{Joint and Factor-Level Results}
	
	Table~\ref{tab:span-joint} reports the joint tests. At the daily frequency, both spanning nulls are strongly rejected, so neither model fully absorbs the other's unique factors. The economic magnitudes are nevertheless asymmetric. Adding q5-specific factors to FF6 raises the annualized maximum Sharpe ratio from 1.311 to 2.205, with
	$\Delta\operatorname{SR}^{2}=3.143$. Adding FF6-specific factors to q5 raises it from 2.077 to 2.205, with
	$\Delta\operatorname{SR}^{2}=0.549$.
	
	\FloatBarrier
	\begin{table}[H]
		\centering
		\onehalfspacing
		\caption{Joint Factor-Model Spanning Tests and Maximum Sharpe Ratios}
		\label{tab:span-joint}
		\small
		\setlength{\tabcolsep}{4.2pt}
		\begin{tabular}{llrrrrrr}
			\toprule
			Frequency & Null hypothesis & GRS & GRS $p$ & HAC Wald & HAC $p$
			& $\operatorname{SR}_{0}$ & $\Delta\operatorname{SR}^{2}$ \\
			\midrule
			Daily   & FF6 $\supseteq$ q5 & 45.197 & $<0.001$ & 148.177 & $<0.001$ & 1.311 & 3.143 \\
			Daily   & q5 $\supseteq$ FF6 &  6.248 & $<0.001$ &  23.086 & $<0.001$ & 2.077 & 0.549 \\
			Monthly & FF6 $\supseteq$ q5 & 34.500 & $<0.001$ & 114.540 & $<0.001$ & 1.222 & 2.687 \\
			Monthly & q5 $\supseteq$ FF6 &  2.217 & 0.051 &  11.426 & 0.044 & 1.982 & 0.255 \\
			\bottomrule
		\end{tabular}
		
		\vspace{0.4em}
		\parbox{0.96\textwidth}{\footnotesize
			Notes: $FF6\supseteq q5$ denotes the null that FF6 jointly spans ME, IA, ROE, and EG. $q5\supseteq FF6$ denotes the null that q5 jointly spans SMB, HML, RMW, CMA, and MOM. $\operatorname{SR}_{0}$ is the annualized sample maximum Sharpe ratio of the benchmark model, and $\Delta\operatorname{SR}^{2}$ is the annualized increase from adding the excluded block. The table uses the Fama--French market factor; the global--q market factor produces the same conclusions. HAC lags are 21 trading days and six months.}
	\end{table}
	\FloatBarrier
	
	The monthly results preserve the asymmetry. The null that FF6 spans the q5-specific factors is strongly rejected. In the reverse direction, the GRS and HAC $p$-values are 0.051 and 0.044, respectively, so the conclusion at the 5\% level depends on the inference method. Taken together with the daily results, the evidence does not support robust spanning in either direction. It does show that q5 has the higher standalone maximum Sharpe ratio and that the q5-specific factor block contributes substantially more incremental maximum-Sharpe information to FF6 than the FF6-specific block contributes to q5.
	
	Table~\ref{tab:span-factor-level} identifies the source of this asymmetry. When the q5-specific factors are regressed on FF6, ME is almost fully explained and IA is insignificant at the 5\% level. ROE and EG, however, retain annualized alphas of 313.6 bp and 864.4 bp. EG alone contributes an individual
	$\Delta\operatorname{SR}^{2}$ of 2.425 and is the main source of the incremental maximum-Sharpe contribution of the q5-specific factor block.
	
	In the reverse regressions, q5 leaves insignificant alphas for HML, RMW, CMA, and MOM. SMB remains significant, with an annualized alpha of 92.3 bp, and is the principal source of the marginal monthly rejection. The individual
	$\Delta\operatorname{SR}^{2}$ values should not be summed because the joint statistic uses the full residual covariance matrix.
	
	\FloatBarrier
	\begin{table}[H]
		\centering
		\onehalfspacing
		\caption{Monthly Factor-Level Spanning Regressions}
		\label{tab:span-factor-level}
		\small
		\setlength{\tabcolsep}{5.2pt}
		\begin{tabular}{llrrrr}
			\toprule
			Benchmark & Target factor & Annualized alpha (bp) & HAC $p$ & $R^{2}$
			& Individual $\Delta\operatorname{SR}^{2}$ \\
			\midrule
			\multicolumn{6}{l}{\textit{Panel A: Pricing q5-Specific Factors with FF6}} \\
			FF6 & ME  &  17.6 & 0.605 & 0.950 & 0.005 \\
			FF6 & IA  &  74.6 & 0.070 & 0.855 & 0.074 \\
			FF6 & ROE & 313.6 & $<0.001$ & 0.642 & 0.347 \\
			FF6 & EG  & 864.4 & $<0.001$ & 0.449 & 2.425 \\
			\addlinespace
			\multicolumn{6}{l}{\textit{Panel B: Pricing FF6-Specific Factors with q5}} \\
			q5 & SMB &  92.3 & 0.006 & 0.956 & 0.170 \\
			q5 & HML & 119.8 & 0.348 & 0.485 & 0.025 \\
			q5 & RMW &  13.3 & 0.895 & 0.473 & 0.001 \\
			q5 & CMA & -13.7 & 0.755 & 0.861 & 0.003 \\
			q5 & MOM & -42.0 & 0.843 & 0.277 & 0.001 \\
			\bottomrule
		\end{tabular}
		
		\vspace{0.4em}
		\parbox{0.96\textwidth}{\footnotesize
			Notes: The sample contains 696 monthly observations from January 1967 through December 2024. Alphas are monthly intercepts multiplied by 12 and reported in basis points. HAC $p$-values use six Newey--West lags. Individual $\Delta\operatorname{SR}^{2}$ is the annualized increase obtained by adding the target factor alone to the benchmark model.}
	\end{table}
	\FloatBarrier
	
	\subsection{Relation to the Random-Portfolio Analysis}
	
	The spanning results delimit the interpretation of the random-portfolio evidence. Under the maximum-Sharpe criterion, q5 offers the larger in-sample opportunity set, although neither model robustly spans the other in every specification. Smaller average absolute alphas for FF6 or another model on a random-portfolio population therefore do not establish universal superiority over q5.
	
	The converse also holds. A larger maximum Sharpe ratio need not imply smaller return-unit errors on external portfolios. Spanning tests ask how much the opportunity set expands when excluded-factor residuals are optimally combined. Random-portfolio tests ask how large and stable the pricing errors are for individual portfolios generated by a prespecified rule. The two objectives may yield different rankings without contradiction.
	
	\section{Home-Field Bias in Factor-Model Performance}
	\label{sec:homebias}
	
	This section compares factor-model performance across three conventional test-asset families: 367 Fama--French characteristic-sorted portfolios, 1,202 global--q anomaly portfolios, and 47 industry portfolios. All series have complete daily returns over the common sample from January 3, 1967, through December 31, 2024. The balanced-panel restriction removes sample-coverage differences, but limits the analysis to long-lived portfolios.
	
	Home-field bias refers to systematic rank shifts when models are evaluated on test assets close to the characteristic space emphasized in their construction or validation. It does not require a model to rank first on every familiar test set.
	
	\subsection{Model Rankings across Test-Asset Families}
	
	Table~\ref{tab:homebias-model} reports mean absolute pricing errors and significant-alpha shares.
	
	\FloatBarrier
	\begin{table}[H]
		\centering
		\onehalfspacing
		\caption{Factor-Model Performance on Conventional Test Assets}
		\label{tab:homebias-model}
		\small
		\setlength{\tabcolsep}{6pt}
		
		\begin{tabular}{lrrrrrr}
			\toprule
			Test assets
			& CAPM & FF3 & Carhart & FF5 & FF6 & q5 \\
			\midrule
			
			\multicolumn{7}{l}{
				\textit{Panel A: Mean Absolute Pricing Error (bp)}
			} \\
			
			Fama--French characteristic sorts
			& 243.4
			& 145.5
			& 137.0
			& 126.8
			& \textbf{121.2}
			& 187.6 \\
			
			global--q anomalies
			& 151.2
			& 154.0
			& 124.4
			& 133.9
			& \textbf{113.6}
			& 114.9 \\
			
			Industry portfolios
			& \textbf{178.5}
			& 238.2
			& 203.8
			& 275.0
			& 247.9
			& 181.3 \\
			
			\midrule
			
			\multicolumn{7}{l}{
				\textit{Panel B: Share of Significant Alphas (\%)}
			} \\
			
			Fama--French characteristic sorts
			& 36.0
			& 25.6
			& 22.3
			& 16.9
			& \textbf{15.8}
			& 33.8 \\
			
			global--q anomalies
			& 31.9
			& 36.3
			& 27.0
			& 30.8
			& 23.8
			& \textbf{20.4} \\
			
			Industry portfolios
			& 8.5
			& 29.8
			& 12.8
			& 31.9
			& 34.0
			& \textbf{4.3} \\
			
			\bottomrule
		\end{tabular}
		
		\vspace{0.4em}
		\begin{minipage}{0.95\textwidth}
			\footnotesize
			\textit{Notes:} The Fama--French characteristic-sorted, global--q anomaly,
			and industry test sets contain 367, 1,202, and 47 portfolios, respectively.
			Panel A reports the cross-sectional mean of annualized absolute pricing
			errors in basis points. Panel B reports the percentage of portfolios with
			an alpha significant at the 5\% level using Newey--West standard errors.
			Boldface denotes the minimum in each row.
		\end{minipage}
	\end{table}
	\FloatBarrier
	
	On Fama--French characteristic-sorted portfolios, FF6 achieves the lowest mean absolute pricing error and the lowest share of significant alphas. The other Fama--French models also record lower mean absolute pricing errors than q5. In matched portfolio comparisons, FF6 produces a smaller absolute pricing error than q5 for 67.8\% of the portfolios.
	
	On global--q anomalies, q5 improves sharply: it outperforms FF3 on both reported metrics and on 61.1\% of matched absolute errors. FF6 retains a slightly lower mean absolute error, but q5 has the lower median error and significant-alpha share. Industry portfolios produce a third ordering: the CAPM has the smallest mean absolute error, while q5 has the lowest significant-alpha share.
	
	Thus, home-field bias is a rank-shift phenomenon rather than a claim that each model must dominate its familiar test assets. Fama--French models perform broadly well on Fama--French sorts, q5 improves on global--q anomalies, and industry portfolios generate a different ranking from either characteristic-based family.
	
	\subsection{Conditional Contributions of the q5 Factors}
	
	To examine the source of this dependence, I compare q3, which contains the market, size, and investment factors, with q4--ROE, q4--EG, and the full q5 model.
	
	\FloatBarrier
	\begin{table}[H]
		\centering
		\onehalfspacing
		\caption{q5 Factor-Ablation Results across Test-Asset Families}
		\label{tab:homebias-ablation}
		\small
		\setlength{\tabcolsep}{8pt}
		
		\begin{tabular}{lrrrr}
			\toprule
			Test assets
			& q3
			& q4--ROE
			& q4--EG
			& q5 \\
			\midrule
			
			\multicolumn{5}{l}{
				\textit{Panel A: Mean Absolute Pricing Error (bp)}
			} \\
			
			Fama--French characteristic sorts
			& \textbf{138.2}
			& 143.7
			& 178.4
			& 187.6 \\
			
			global--q anomalies
			& 133.2
			& \textbf{105.9}
			& 108.9
			& 114.9 \\
			
			Industry portfolios
			& 240.9
			& 204.8
			& 186.7
			& \textbf{181.3} \\
			
			\midrule
			
			\multicolumn{5}{l}{
				\textit{Panel B: Share of Significant Alphas (\%)}
			} \\
			
			Fama--French characteristic sorts
			& 21.0
			& \textbf{19.3}
			& 31.3
			& 33.8 \\
			
			global--q anomalies
			& 27.9
			& \textbf{17.3}
			& 18.1
			& 20.4 \\
			
			Industry portfolios
			& 21.3
			& 12.8
			& 6.4
			& \textbf{4.3} \\
			
			\bottomrule
		\end{tabular}
		
		\vspace{0.4em}
		\begin{minipage}{0.95\textwidth}
			\footnotesize
			\textit{Notes:} q3 contains the market, size, and investment factors.
			q4--ROE and q4--EG add the return-on-equity and expected-growth factors,
			respectively, to q3. q5 contains both factors. Panel A reports the
			cross-sectional mean of annualized absolute pricing errors in basis points.
			Panel B reports the percentage of portfolios with an alpha significant at
			the 5\% level using Newey--West standard errors. Boldface denotes the
			minimum in each row.
		\end{minipage}
	\end{table}
	\FloatBarrier
	
	On Fama--French characteristic sorts, q3 has the smallest mean absolute error. Adding ROE slightly increases it, while adding EG produces substantially larger errors and higher significant-alpha shares. On global--q anomalies, both ROE and EG improve on q3, with q4--ROE performing best. On industry portfolios, the additions improve fit sequentially and the full q5 performs best.
	
	The same q5 factors therefore have different marginal effects across test-asset families. This also separates portfolio-level fit from the spanning results in \cref{sec:span}: EG substantially expands q5's sample maximum Sharpe ratio, yet increases return-unit pricing errors on Fama--French sorts while reducing them on global--q anomalies and industry portfolios.
	
	The three test-asset families therefore produce neither a construction-invariant model ranking nor construction-invariant factor contributions. In the notation of equation~\eqref{eq:distribution_loss_lit}, changing the test-asset family changes $Q_{\mathcal{D}}$ and emphasizes different directions of the pricing-error vector. The next section extends this static comparison to random portfolios with alternative stock-selection, initial-weighting, holding, and rebalancing rules.
	
	\section{Construction Dependence in Random Portfolios}
	\label{sec:random}
	
	This section extends the analysis from conventional characteristic-sorted portfolios to random portfolios formed without ex ante sorting on firm characteristics. The baseline design fixes the stock-selection ratio at 5\% and combines two sampling rules, MEP and UNIF, with four initial weighting schemes, EW, VW, SQME, and RAND. Each of the eight cells contains 500 portfolios.
	
	I apply two weight-management rules to the same selected stocks and initial weights. Buy and hold (BH) allows weights to drift with realized returns, whereas constant weight (CW) restores the initial target weights each day. The paired design therefore isolates the effect of post-formation weight management. Model fit is evaluated using annualized pricing errors, absolute pricing errors, tail errors, significant-alpha shares, and $R^2$. Because the random portfolios overlap substantially, the significant-alpha share is interpreted descriptively rather than as a collection of independent tests.
	
	\subsection{Buy and Hold}
	
	Under BH, winners gain weight and losers lose weight over the holding period. Pricing errors therefore reflect both the initial construction and the subsequent path of relative returns.
	
\begin{figure}[p]
		\centering
		
		\begin{subfigure}[t]{\linewidth}
			\centering
			\includegraphics[width=\linewidth]{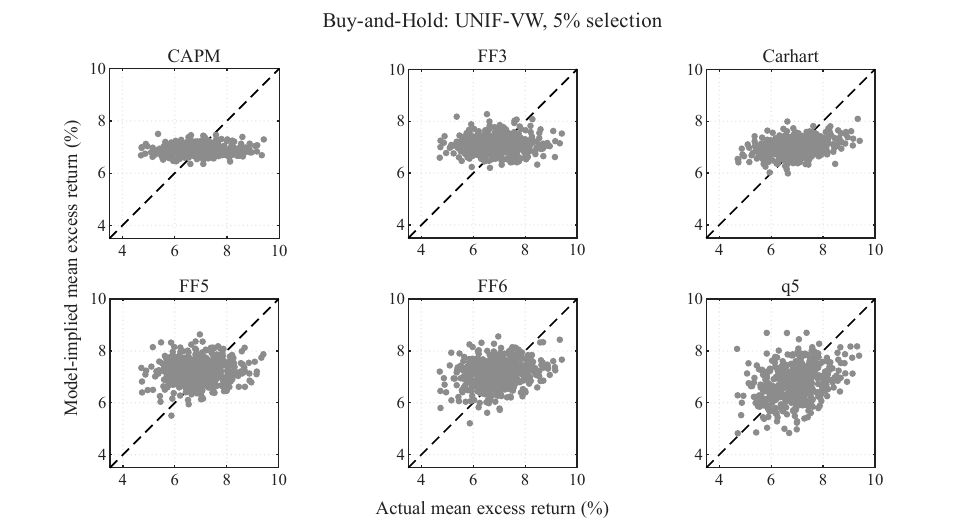}
			\caption{UNIF--VW}
			\label{fig:bh-unif-vw}
		\end{subfigure}
		
		\vspace{0.6em}
		
		\begin{subfigure}[t]{\linewidth}
			\centering
			\includegraphics[width=\linewidth]{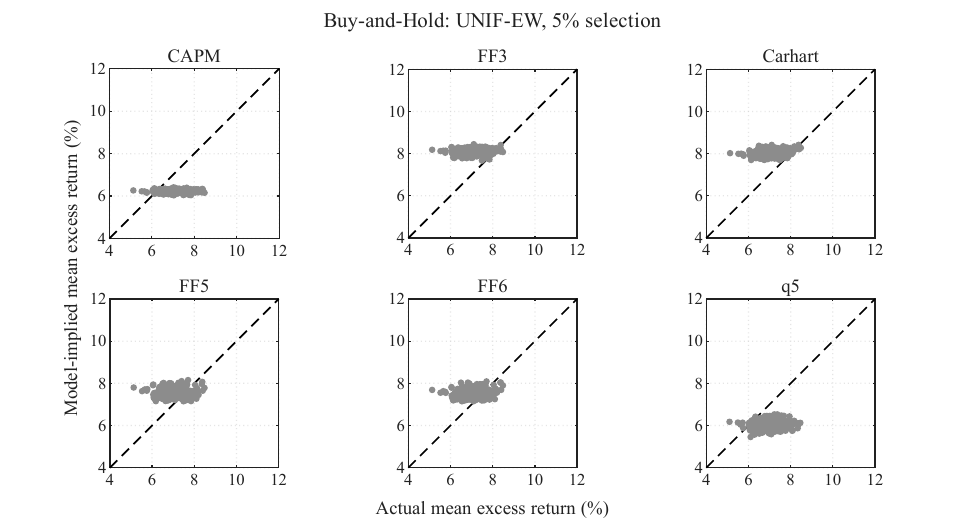}
			\caption{UNIF--EW}
			\label{fig:bh-unif-ew}
		\end{subfigure}
		
		\caption{Realized and Model-Implied Mean Excess Returns:
			Buy and Hold, UNIF, 5\% Selection Ratio}
		\label{fig:bh-unif-fit}
		
		\vspace{0.35em}
		\begin{minipage}{0.94\linewidth}
			\footnotesize
			\textit{Notes:} Each panel contains 500 random portfolios. The horizontal
			axis reports the realized annualized mean excess return. The vertical axis
			reports the model-implied annualized mean excess return, computed from
			estimated factor loadings and sample mean factor returns. The dashed line
			is the 45-degree line. Points below the line have positive estimated alphas.
		\end{minipage}
	\end{figure}

	Figure~\ref{fig:bh-unif-fit} illustrates this point for two UNIF designs. Each point is one random portfolio. The horizontal axis reports the realized annualized mean excess return, and the vertical axis reports the model-implied mean excess return. Points below the 45-degree line have positive alphas.
	
	\FloatBarrier
	\FloatBarrier
	
	The two panels produce different model rankings even though they use the same UNIF stock draws. Under VW, Carhart lies closest to the 45-degree line and has the smallest mean absolute pricing error. Under EW, FF5 and FF6 are closest, with FF6 performing best. Initial weighting therefore changes the cross-sectional return range, the level of model-implied returns, and the resulting model ranking.
	
	Table~\ref{tab:bh-cell-main} reports all eight BH cells.
	
	\FloatBarrier
	\begin{table}[H]
		\centering
		\onehalfspacing
		\caption{Mean Absolute Pricing Errors and Significant-Alpha Shares:
			Buy and Hold, 5\% Selection Ratio}
		\label{tab:bh-cell-main}
		\small
		\setlength{\tabcolsep}{5.5pt}
		
		\begin{tabular}{llrrrrrr}
			\toprule
			Sampling & Initial weight
			& CAPM & FF3 & Carhart & FF5 & FF6 & q5 \\
			\midrule
			
			\multicolumn{8}{l}{
				\textit{Panel A: Mean Absolute Pricing Error (bp)}
			} \\
			
			MEP  & EW
			& \textbf{26.7}
			& 32.8
			& 31.1
			& 29.5
			& 27.7
			& 56.1 \\
			
			MEP  & VW
			& 23.6
			& 46.8
			& 50.5
			& 10.7
			& \textbf{10.1}
			& 110.3 \\
			
			MEP  & SQME
			& \textbf{14.3}
			& 15.6
			& 15.0
			& 20.8
			& 17.6
			& 20.3 \\
			
			MEP  & RAND
			& \textbf{30.8}
			& 36.6
			& 34.5
			& 33.9
			& 31.9
			& 56.8 \\
			
			\addlinespace
			
			UNIF & EW
			& 80.0
			& 107.1
			& 102.2
			& 61.5
			& \textbf{59.3}
			& 94.8 \\
			
			UNIF & VW
			& 67.6
			& 72.1
			& \textbf{62.7}
			& 76.3
			& 70.5
			& 73.3 \\
			
			UNIF & SQME
			& 52.0
			& 66.5
			& 62.5
			& 48.2
			& \textbf{45.4}
			& 71.0 \\
			
			UNIF & RAND
			& 82.3
			& 107.8
			& 102.7
			& 64.6
			& \textbf{62.1}
			& 97.3 \\
			
			\midrule
			
			\multicolumn{8}{l}{
				\textit{Panel B: Share of Significant Alphas (\%)}
			} \\
			
			MEP  & EW
			& \textbf{2.4}
			& 8.8
			& 6.0
			& 6.0
			& 3.6
			& 34.0 \\
			
			MEP  & VW
			& \textbf{0.0}
			& 14.4
			& 16.4
			& \textbf{0.0}
			& \textbf{0.0}
			& 100.0 \\
			
			MEP  & SQME
			& \textbf{0.4}
			& 3.8
			& 2.6
			& 8.6
			& 3.2
			& 4.8 \\
			
			MEP  & RAND
			& \textbf{2.6}
			& 10.0
			& 6.0
			& 6.6
			& 5.6
			& 26.4 \\
			
			\addlinespace
			
			UNIF & EW
			& \textbf{0.0}
			& 47.6
			& 44.0
			& 14.6
			& 11.8
			& 29.0 \\
			
			UNIF & VW
			& 5.0
			& 6.4
			& \textbf{3.2}
			& 8.0
			& 5.6
			& 4.6 \\
			
			UNIF & SQME
			& \textbf{1.2}
			& 23.6
			& 18.8
			& 12.4
			& 9.4
			& 25.4 \\
			
			UNIF & RAND
			& \textbf{0.2}
			& 38.6
			& 33.8
			& 11.6
			& 11.0
			& 26.0 \\
			
			\bottomrule
		\end{tabular}
		
		\vspace{0.4em}
		\begin{minipage}{0.95\textwidth}
			\footnotesize
			\textit{Notes:} Each cell contains 500 BH portfolios. Panel A reports
			the cross-sectional mean of annualized absolute pricing errors in basis
			points. Panel B reports the percentage of portfolios with an alpha
			significant at the 5\% level using Newey--West standard errors.
			Boldface denotes the minimum in each row.
		\end{minipage}
	\end{table}
	\FloatBarrier
	
	Under MEP, the CAPM performs best for EW, SQME, and RAND, whereas FF6 performs best for VW. Under UNIF, FF6 performs best for EW, SQME, and RAND, and Carhart performs best for VW. q5 never attains the smallest mean absolute pricing error in the eight BH cells. Its weakest result occurs in MEP--VW, where the mean absolute error reaches 110.3 bp and all alphas are significant.
	
	The BH evidence therefore does not produce a single model ranking. CAPM, Carhart, and FF6 each perform best in at least one design. The important result is that rankings and error distributions move with both the sampling rule and the initial weights.
	
	\subsection{Constant Weight}
	
	Under CW, the portfolio is rebalanced daily to its formation weights. This removes cumulative weight drift but creates a repeated sell-winners, buy-losers mechanism. The resulting return process differs sharply from BH.
	
\begin{figure}[p]
		\centering
		
		\begin{subfigure}[t]{\linewidth}
			\centering
			\includegraphics[width=0.985\linewidth]{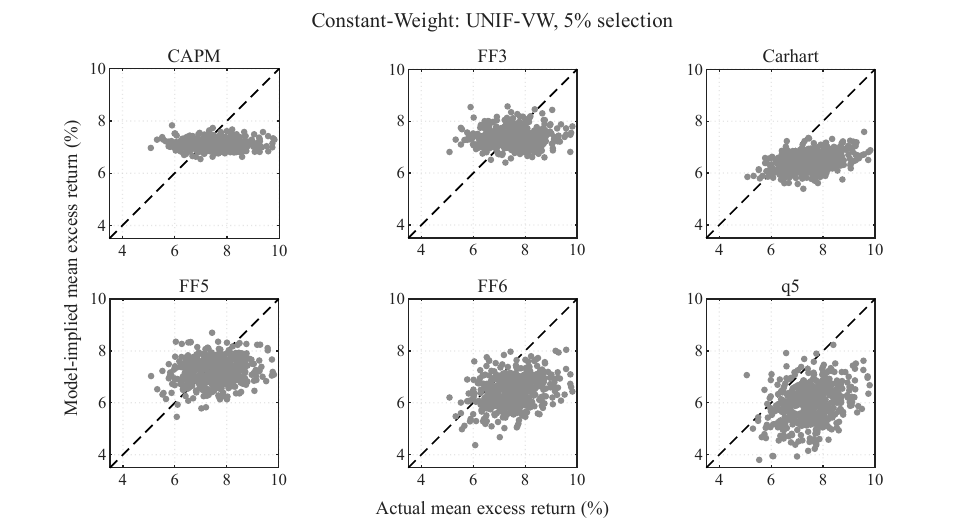}
			\caption{UNIF--VW}
			\label{fig:cw-unif-vw}
		\end{subfigure}
		
		\vspace{0.6em}
		
		\begin{subfigure}[t]{\linewidth}
			\centering
			\includegraphics[width=0.985\linewidth]{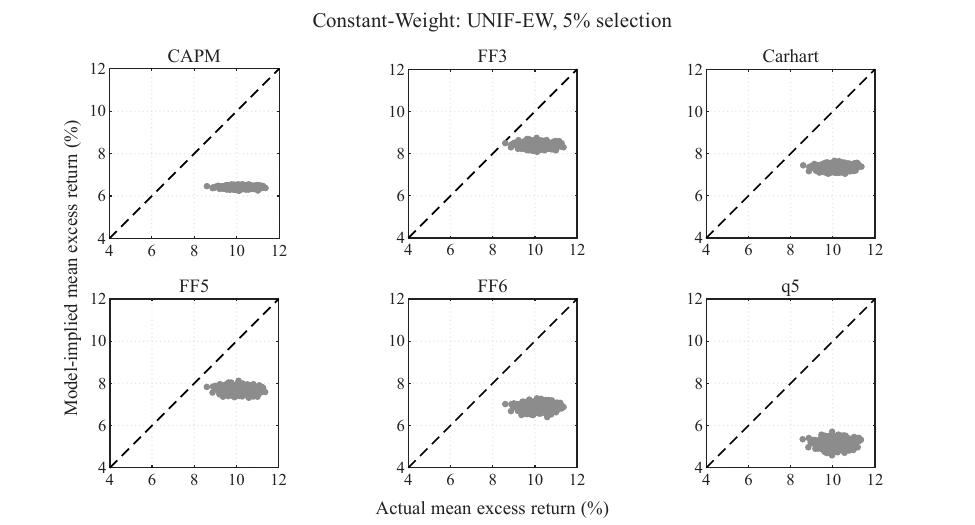}
			\caption{UNIF--EW}
			\label{fig:cw-unif-ew}
		\end{subfigure}
		
		\caption{Realized and Model-Implied Mean Excess Returns:
			Constant Weight, UNIF, 5\% Selection Ratio}
		\label{fig:cw-unif-fit}
		
		\vspace{0.35em}
		\begin{minipage}{0.94\linewidth}
			\footnotesize
			\textit{Notes:} Each panel contains 500 random portfolios. The axes and
			the 45-degree line are defined as in Figure~\ref{fig:bh-unif-fit}.
		\end{minipage}
	\end{figure}

	Figure~\ref{fig:cw-unif-fit} again compares UNIF--VW and UNIF--EW.
	
	\FloatBarrier
	\FloatBarrier
	
	In UNIF--VW, the models compress the cross-section of realized mean returns. FF3 lies closest to the 45-degree line, whereas q5 leaves especially large positive alphas among high-return portfolios. In UNIF--EW, the portfolios cluster in a narrow high-return range and all models lie below the 45-degree line. The dominant failure is therefore a level effect rather than a failure to rank portfolios. FF3 remains closest to the line, while q5 produces the largest underprediction.
	
	Table~\ref{tab:cw-cell-main} summarizes all eight CW cells.
	
	\FloatBarrier
	\begin{table}[H]
		\centering
		\onehalfspacing
		\caption{Mean Absolute Pricing Errors and Significant-Alpha Shares:
			Constant Weight, 5\% Selection Ratio}
		\label{tab:cw-cell-main}
		\small
		\setlength{\tabcolsep}{5.5pt}
		
		\begin{tabular}{llrrrrrr}
			\toprule
			Sampling & Initial weight
			& CAPM & FF3 & Carhart & FF5 & FF6 & q5 \\
			\midrule
			
			\multicolumn{8}{l}{
				\textit{Panel A: Mean Absolute Pricing Error (bp)}
			} \\
			
			MEP  & EW
			& 64.2
			& \textbf{23.0}
			& 103.5
			& 45.8
			& 119.1
			& 212.5 \\
			
			MEP  & VW
			& \textbf{10.5}
			& 67.9
			& 145.3
			& 46.8
			& 110.6
			& 14.7 \\
			
			MEP  & SQME
			& \textbf{24.3}
			& 33.1
			& 117.8
			& 36.3
			& 104.2
			& 117.7 \\
			
			MEP  & RAND
			& 64.0
			& \textbf{27.8}
			& 103.4
			& 47.5
			& 119.2
			& 212.2 \\
			
			\addlinespace
			
			UNIF & EW
			& 368.3
			& \textbf{169.1}
			& 276.4
			& 241.8
			& 324.2
			& 490.3 \\
			
			UNIF & VW
			& 78.3
			& \textbf{72.7}
			& 114.0
			& 77.6
			& 115.9
			& 158.2 \\
			
			UNIF & SQME
			& 153.9
			& \textbf{46.4}
			& 137.6
			& 91.0
			& 169.4
			& 288.3 \\
			
			UNIF & RAND
			& 367.5
			& \textbf{168.3}
			& 275.7
			& 241.1
			& 323.6
			& 490.1 \\
			
			\midrule
			
			\multicolumn{8}{l}{
				\textit{Panel B: Share of Significant Alphas (\%)}
			} \\
			
			MEP  & EW
			& 13.8
			& \textbf{0.2}
			& 79.6
			& 5.6
			& 93.4
			& 100.0 \\
			
			MEP  & VW
			& \textbf{0.0}
			& 29.4
			& 100.0
			& 1.6
			& 100.0
			& \textbf{0.0} \\
			
			MEP  & SQME
			& \textbf{0.0}
			& 4.8
			& 99.8
			& 4.6
			& 99.4
			& 99.8 \\
			
			MEP  & RAND
			& 14.0
			& \textbf{0.4}
			& 66.6
			& 8.8
			& 84.0
			& 100.0 \\
			
			\addlinespace
			
			UNIF & EW
			& 99.8
			& \textbf{77.8}
			& 99.8
			& 98.6
			& 100.0
			& 100.0 \\
			
			UNIF & VW
			& 6.0
			& \textbf{5.0}
			& 18.8
			& 6.6
			& 19.2
			& 37.0 \\
			
			UNIF & SQME
			& 48.8
			& \textbf{3.8}
			& 76.4
			& 32.4
			& 90.6
			& 100.0 \\
			
			UNIF & RAND
			& 98.8
			& \textbf{67.0}
			& 99.4
			& 97.4
			& 100.0
			& 100.0 \\
			
			\bottomrule
		\end{tabular}
		
		\vspace{0.4em}
		\begin{minipage}{0.95\textwidth}
			\footnotesize
			\textit{Notes:} Each cell contains 500 CW portfolios. Panel A reports
			the cross-sectional mean of annualized absolute pricing errors in basis
			points. Panel B reports the percentage of portfolios with an alpha
			significant at the 5\% level using Newey--West standard errors.
			Boldface denotes the minimum in each row.
		\end{minipage}
	\end{table}
	\FloatBarrier
	
	FF3 has the smallest mean absolute pricing error in six of the eight CW cells. The exceptions are MEP--VW and MEP--SQME, where the CAPM performs best. MEP--VW is also an important exception for q5: its mean absolute error is only 14.7 bp, close to the CAPM's 10.5 bp, and none of its alphas is significant.
	
	Outside that cell, q5 errors rise sharply. Mean absolute errors exceed 200 bp in MEP--EW and MEP--RAND and approach 490 bp in UNIF--EW and UNIF--RAND. Carhart and FF6 also perform poorly in several CW cells, often with widespread significant positive alphas. These results are not explained by large differences in average $R^2$. Daily rebalancing changes the return process in ways that materially affect intercepts even when time-series fit remains high.
	
	\subsection{Paired Comparison of BH and CW}
	
	Because BH and CW use the same stocks, initial weights, constituent returns, and delisting treatment, their paired difference isolates post-formation weight management. Table~\ref{tab:bh-cw-pair-main} reports the direct comparison.
	
	\FloatBarrier
	\begin{table}[H]
		\centering
		\onehalfspacing
		\caption{Paired Comparison of BH and CW: 5\% Selection Ratio}
		\label{tab:bh-cw-pair-main}
		\small
		\setlength{\tabcolsep}{5.5pt}
		
		\begin{tabular}{llrrrrrr}
			\toprule
			Sampling & Initial weight
			& CAPM & FF3 & Carhart & FF5 & FF6 & q5 \\
			\midrule
			
			\multicolumn{8}{l}{
				\textit{Panel A: Mean BH--CW Difference in Absolute Pricing Error (bp)}
			} \\
			
			MEP  & EW
			& $-37.5$
			& \textbf{$9.7$}
			& $-72.4$
			& $-16.3$
			& $-91.4$
			& $-156.4$ \\
			
			MEP  & VW
			& \textbf{$13.2$}
			& $-21.1$
			& $-94.8$
			& $-36.1$
			& $-100.5$
			& \textbf{$95.6$} \\
			
			MEP  & SQME
			& $-10.0$
			& $-17.4$
			& $-102.8$
			& $-15.5$
			& $-86.7$
			& $-97.4$ \\
			
			MEP  & RAND
			& $-33.3$
			& \textbf{$8.8$}
			& $-68.9$
			& $-13.7$
			& $-87.3$
			& $-155.5$ \\
			
			\addlinespace
			
			UNIF & EW
			& $-288.3$
			& $-62.0$
			& $-174.2$
			& $-180.3$
			& $-264.9$
			& $-395.5$ \\
			
			UNIF & VW
			& $-10.8$
			& $-0.6$
			& $-51.3$
			& $-1.3$
			& $-45.5$
			& $-84.9$ \\
			
			UNIF & SQME
			& $-101.9$
			& \textbf{$20.1$}
			& $-75.1$
			& $-42.8$
			& $-124.0$
			& $-217.4$ \\
			
			UNIF & RAND
			& $-285.1$
			& $-60.5$
			& $-173.0$
			& $-176.5$
			& $-261.5$
			& $-392.8$ \\
			
			\midrule
			
			\multicolumn{8}{l}{
				\textit{Panel B: Share of Pairs with Smaller BH Absolute Alpha (\%)}
			} \\
			
			MEP  & EW
			& 95.0
			& \textbf{35.2}
			& 94.6
			& 64.0
			& 98.4
			& 100.0 \\
			
			MEP  & VW
			& \textbf{20.6}
			& 100.0
			& 100.0
			& 94.6
			& 100.0
			& \textbf{0.0} \\
			
			MEP  & SQME
			& 69.2
			& 89.6
			& 100.0
			& 70.6
			& 99.6
			& 99.8 \\
			
			MEP  & RAND
			& 88.8
			& \textbf{36.8}
			& 89.6
			& 62.8
			& 93.8
			& 100.0 \\
			
			\addlinespace
			
			UNIF & EW
			& 100.0
			& 73.2
			& 98.4
			& 98.4
			& 99.6
			& 100.0 \\
			
			UNIF & VW
			& 62.6
			& 52.6
			& 74.2
			& 53.0
			& 69.4
			& 81.4 \\
			
			UNIF & SQME
			& 99.2
			& \textbf{37.2}
			& 82.0
			& 71.8
			& 92.2
			& 100.0 \\
			
			UNIF & RAND
			& 100.0
			& 71.8
			& 96.4
			& 97.2
			& 99.0
			& 100.0 \\
			
			\bottomrule
		\end{tabular}
		
		\vspace{0.4em}
		\begin{minipage}{0.95\textwidth}
			\footnotesize
			\textit{Notes:} Panel A reports the mean of
			$|\alpha^{BH}|-|\alpha^{CW}|$ across 500 matched portfolios. Negative
			values indicate smaller BH errors. Panel B reports the percentage of pairs
			for which $|\alpha^{BH}|<|\alpha^{CW}|$. Boldface identifies cells in
			which CW performs better both on average and for a majority of pairs.
		\end{minipage}
	\end{table}
	\FloatBarrier
	
	BH generally produces smaller pricing errors. Carhart, FF5, and FF6 favor BH in all eight cells. CAPM and q5 favor BH in seven cells. The BH--CW differences reach several hundred basis points in UNIF--EW and UNIF--RAND.
	
	MEP--VW is the clearest exception. For q5, CW reduces mean absolute pricing error by 95.6 bp, and CW outperforms BH for all 500 matched portfolios. CAPM also favors CW in this cell. The weak CW performance of q5 elsewhere therefore cannot be attributed to daily rebalancing alone. Its effect depends on the interaction between sampling and initial weighting.
	
	FF3 is the least directional model in the paired comparison. CW performs better in MEP--EW, MEP--RAND, and UNIF--SQME, whereas BH performs better in several other cells. UNIF--VW is close to even. This mixed pattern is consistent with FF3's relatively stable performance across construction rules.
	
	\subsection{Summary}
	
	Random-portfolio model evaluation depends on both static construction and dynamic weight management. BH favors FF5 and FF6 in many designs, whereas CW generally favors FF3. Carhart and FF6 perform well under BH but leave broad positive intercepts under daily rebalancing.
	
	q5 is not worst in every individual design. In particular, it performs well under MEP--VW with CW. Across the eight cells, however, its mean absolute pricing errors are the largest under both BH and CW, and its BH--CW changes are also the most pronounced.
	
	These results do not overturn the spanning evidence or provide a universal ranking of factor models. Spanning tests evaluate relative mean--variance opportunity sets. The analysis here evaluates return-unit pricing errors on portfolios drawn from a prespecified asset-generating distribution. The main conclusion is that test-asset construction, including post-formation weight management, is an economically important part of factor-model evaluation.
	
	\section{q5 Factor-Ablation Tests}
	\label{sec:ablation}
	
	The preceding results show that q5 pricing errors vary substantially with test-asset construction and weight management. This section decomposes that variation by factor block. The objective is not to reassess the economic validity of the individual factors. It is to examine how portfolio-level pricing errors change when ROE and EG are added to a common q3 benchmark.
	
	Define
	\begin{align}
		q3
		&=
		\left\{
		MKT,\,
		ME,\,
		IA
		\right\}, \\
		q4
		&=
		q3 \cup
		\left\{
		ROE
		\right\}, \\
		q3{+}EG
		&=
		q3 \cup
		\left\{
		EG
		\right\}, \\
		q5
		&=
		q3 \cup
		\left\{
		ROE,\,
		EG
		\right\}.
	\end{align}
	
	The main outcome is annualized mean absolute pricing error. Relative to q3, the change for model $M$ is
	\begin{equation}
		\Delta \left|\alpha\right|_{M}
		=
		\overline{\left|\alpha^{M}\right|}
		-
		\overline{\left|\alpha^{q3}\right|}.
	\end{equation}
	Negative values indicate that the added factor reduces pricing errors. Positive values indicate deterioration. This statistic evaluates return-unit errors on a prespecified test-asset population rather than changes in the maximum Sharpe ratio.
	
	\subsection{Conventional Test Portfolios}
	
	Table~\ref{tab:ablation-conventional} compares the four models on 367 Fama--French characteristic-sorted portfolios, 47 industry portfolios, and 1,202 global--q anomaly portfolios.
	
	\FloatBarrier
	\begin{table}[htbp]
		\centering
		\onehalfspacing
		\caption{q5 Factor-Ablation Tests on Conventional Test Portfolios}
		\label{tab:ablation-conventional}
		\small
		\begin{tabular}{lrrrrr}
			\toprule
			Test assets
			& q3
			& q4
			& q3+EG
			& q5
			& $\Delta|{\alpha}|_{q5-q3}$ \\
			\midrule
			Fama--French characteristic sorts
			& \textbf{138.2}
			& 143.7
			& 178.4
			& 187.6
			& 49.4 \\
			Industry portfolios
			& 240.9
			& 204.8
			& 186.7
			& \textbf{181.3}
			& -59.6 \\
			global--q anomalies
			& 133.2
			& \textbf{105.9}
			& 108.9
			& 114.9
			& -18.3 \\
			\bottomrule
		\end{tabular}
		
		\vspace{0.5em}
		\begin{minipage}{0.93\textwidth}
			\footnotesize
			\textit{Notes:} Entries are annualized mean absolute pricing errors in basis points.
			Boldface denotes the smallest value within each test-asset family. The final
			column reports the difference between q5 and q3.
		\end{minipage}
	\end{table}
	\FloatBarrier
	
	On the Fama--French characteristic sorts, q3 performs best. Adding ROE increases mean absolute pricing error by 5.5 bp, while adding EG increases it by 40.2 bp. The full q5 error exceeds the q3 error by 49.4 bp. Higher average $R^2$ therefore does not translate into smaller intercepts.
	
	The industry portfolios produce the opposite result. q5 reduces mean absolute pricing error by 59.6 bp relative to q3 and performs best among the four specifications. On the global--q anomaly portfolios, adding either ROE or EG improves on q3, although the full q5 model performs worse than either one-factor extension.
	
	The marginal contribution of a factor is therefore conditional on the test assets. ROE and EG increase pricing errors on the Fama--French characteristic sorts but reduce them on the industry and global--q portfolios.
	
	\subsection{BH Random Portfolios}
	
	Table~\ref{tab:ablation-random} reports the ablation results for the eight BH designs at the 5\% selection ratio.
	
	\FloatBarrier
	\begin{table}[H]
		\centering
		\onehalfspacing
		\caption{q5 Factor-Ablation Tests on BH Random Portfolios:
			5\% Selection Ratio}
		\label{tab:ablation-random}
		\small
		\setlength{\tabcolsep}{7pt}
		
		\begin{tabular}{llrrrr}
			\toprule
			Sampling & Initial weight
			& q3 & q4 & q3+EG & q5 \\
			\midrule
			
			\multicolumn{6}{l}{
				\textit{Panel A: Mean Absolute Pricing Error (bp)}
			} \\
			
			MEP  & EW
			& 37.8 & \textbf{21.6} & 53.8 & 56.1 \\
			
			MEP  & VW
			& 36.1 & \textbf{28.0} & 103.9 & 110.3 \\
			
			MEP  & SQME
			& \textbf{13.7} & 26.6 & 17.4 & 20.3 \\
			
			MEP  & RAND
			& 41.0 & \textbf{26.1} & 54.5 & 56.8 \\
			
			\addlinespace
			
			UNIF & EW
			& 106.0 & \textbf{50.3} & 72.4 & 94.8 \\
			
			UNIF & VW
			& 71.5 & \textbf{68.7} & 73.4 & 73.3 \\
			
			UNIF & SQME
			& 67.2 & \textbf{39.8} & 60.7 & 71.0 \\
			
			UNIF & RAND
			& 106.4 & \textbf{55.7} & 75.9 & 97.3 \\
			
			\midrule
			
			\multicolumn{6}{l}{
				\textit{Panel B: Share of Significant Alphas (\%)}
			} \\
			
			MEP  & EW
			& 14.0 & \textbf{1.6} & 31.6 & 34.0 \\
			
			MEP  & VW
			& 3.6 & \textbf{0.6} & 100.0 & 100.0 \\
			
			MEP  & SQME
			& \textbf{1.4} & 16.0 & 3.2 & 4.8 \\
			
			MEP  & RAND
			& 12.8 & \textbf{1.8} & 24.8 & 26.4 \\
			
			\addlinespace
			
			UNIF & EW
			& 39.2 & \textbf{7.0} & 14.4 & 29.0 \\
			
			UNIF & VW
			& 6.0 & 4.6 & \textbf{4.4} & 4.6 \\
			
			UNIF & SQME
			& 23.0 & \textbf{4.0} & 18.0 & 25.4 \\
			
			UNIF & RAND
			& 31.4 & \textbf{7.6} & 12.8 & 26.0 \\
			
			\bottomrule
		\end{tabular}
		
		\vspace{0.4em}
		\begin{minipage}{0.95\textwidth}
			\footnotesize
			\textit{Notes:} Each design contains 500 BH portfolios. q3 includes the
			market, size, and investment factors. q4 adds ROE, q3+EG adds EG, and
			q5 includes both. Panel A reports annualized mean absolute pricing errors
			in basis points. Panel B reports the percentage of portfolios with an
			alpha significant at the 5\% level using Newey--West standard errors.
			Boldface denotes the minimum in each row.
		\end{minipage}
	\end{table}
	\FloatBarrier
	
	Adding ROE improves on q3 in seven of the eight BH cells and produces the lowest significant-alpha share in six. MEP--SQME is the only clear exception, with q3 performing best on both metrics.
	
	The contribution of EG is less stable. Adding EG alone improves some UNIF designs but raises pricing errors in all MEP designs and in UNIF--VW. The effect therefore changes with both sampling and initial weighting.
	
	The most consistent pattern appears when EG is added to q4. q5 leaves larger pricing errors than q4 in seven of the eight baseline cells and in 62 of 64 cells across the full selection-ratio grid. The benefit of ROE is therefore typically weakened or reversed once EG is included.
	
	\subsection{CW Random Portfolios}
	
	Table~\ref{tab:ablation-random-cw} applies the same decomposition under daily constant weighting.
	
	\FloatBarrier
	\begin{table}[H]
		\centering
		\onehalfspacing
		\caption{q5 Factor-Ablation Tests on CW Random Portfolios:
			5\% Selection Ratio}
		\label{tab:ablation-random-cw}
		\small
		\setlength{\tabcolsep}{7pt}
		
		\begin{tabular}{llrrrr}
			\toprule
			Sampling & Initial weight
			& q3 & q4 & q3+EG & q5 \\
			\midrule
			
			\multicolumn{6}{l}{
				\textit{Panel A: Mean Absolute Pricing Error (bp)}
			} \\
			
			MEP  & EW
			& \textbf{32.9}
			& 122.7
			& 200.9
			& 212.5 \\
			
			MEP  & VW
			& 81.6
			& 62.7
			& 14.8
			& \textbf{14.7} \\
			
			MEP  & SQME
			& \textbf{46.4}
			& 75.9
			& 114.9
			& 117.7 \\
			
			MEP  & RAND
			& \textbf{36.1}
			& 122.7
			& 200.6
			& 212.2 \\
			
			\addlinespace
			
			UNIF & EW
			& \textbf{198.4}
			& 407.7
			& 455.9
			& 490.3 \\
			
			UNIF & VW
			& \textbf{74.7}
			& 109.0
			& 150.7
			& 158.2 \\
			
			UNIF & SQME
			& \textbf{65.7}
			& 206.5
			& 266.4
			& 288.3 \\
			
			UNIF & RAND
			& \textbf{197.7}
			& 407.1
			& 455.6
			& 490.1 \\
			
			\midrule
			
			\multicolumn{6}{l}{
				\textit{Panel B: Share of Significant Alphas (\%)}
			} \\
			
			MEP  & EW
			& \textbf{1.6}
			& 88.2
			& 100.0
			& 100.0 \\
			
			MEP  & VW
			& 69.4
			& 11.2
			& \textbf{0.0}
			& \textbf{0.0} \\
			
			MEP  & SQME
			& \textbf{12.0}
			& 64.8
			& 99.6
			& 99.8 \\
			
			MEP  & RAND
			& \textbf{1.6}
			& 79.2
			& 99.8
			& 100.0 \\
			
			\addlinespace
			
			UNIF & EW
			& \textbf{86.6}
			& 100.0
			& 100.0
			& 100.0 \\
			
			UNIF & VW
			& \textbf{4.8}
			& 16.0
			& 34.6
			& 37.0 \\
			
			UNIF & SQME
			& \textbf{9.8}
			& 96.2
			& 99.8
			& 100.0 \\
			
			UNIF & RAND
			& \textbf{78.6}
			& 100.0
			& 100.0
			& 100.0 \\
			
			\bottomrule
		\end{tabular}
		
		\vspace{0.4em}
		\begin{minipage}{0.95\textwidth}
			\footnotesize
			\textit{Notes:} Each design contains 500 CW portfolios. Model definitions
			and reported statistics follow Table~\ref{tab:ablation-random}. Boldface
			denotes the minimum in each row.
		\end{minipage}
	\end{table}
	\FloatBarrier
	
	q3 produces the smallest mean absolute pricing error in seven of the eight CW cells. In those cells, adding either ROE or EG increases pricing errors, and q5 is generally the weakest specification.
	
	MEP--VW is the sole major exception. In this cell, q5 and q3+EG produce the smallest errors and no significant alphas. The adverse CW contribution of the added factors is therefore broad but not mechanical.
	
	Across the full 64-cell grid, q3 ranks first in 58 cases. q4, q3+EG, and q5 each exceed the q3 pricing error in 59 cases, while q5 exceeds q4 in 60 cases. Average $R^2$ nevertheless rises when the additional factors are included. Under CW, greater explanatory power for return variation can therefore coexist with larger mean-return pricing errors.
	
	\subsection{BH--CW Comparison}
	
	Because BH and CW use the same stocks and initial weights, their differences reflect post-formation weight management. For q3, the direction of the BH--CW difference varies across cells. Some designs favor BH, others favor CW, and UNIF--VW is close to neutral.
	
	For q5, BH produces smaller errors in seven of the eight cells. The differences are especially large in MEP--EW, MEP--RAND, UNIF--EW, and UNIF--RAND. MEP--VW again reverses the general pattern and favors CW.
	
	q5 pricing errors are therefore more broadly sensitive to weight management than q3 errors, although daily rebalancing is not uniformly harmful. The effect depends on the interaction between the factor set, portfolio composition, and the induced weight path.
	
	\subsection{Summary}
	
	The ablation evidence shows that the contributions of ROE and EG depend on both the test assets and the weight-management rule. On conventional test portfolios, the same factors can either reduce or increase pricing errors. Under BH, adding ROE usually improves q3, but much of that improvement disappears once EG is added. Under CW, q3 is the most stable specification except in MEP--VW.
	
	These findings should not be interpreted as evidence of universal success or failure for any individual factor. They instead show that the marginal contribution of a factor in an incomplete model is conditional on the return directions emphasized by the asset-generating distribution and the portfolio's dynamic weight rule.
	
	\section{Robustness Tests}
	\label{sec:robust}
	
	This section examines whether the baseline findings depend on the investment universe, the treatment of microcaps, portfolio breadth, or the weight-management rule. I first restrict portfolio formation to NYSE common stocks. Subsequent tests exclude stocks below the twentieth percentile of the NYSE market-equity distribution, vary the ALL--UNIF--VW selection ratio from 0.5\% to 100\%, and replace the BH--CW endpoints with intermediate rebalancing frequencies. The objective is not to reproduce a single ranking. It is to determine whether model performance remains sensitive to both static portfolio construction and dynamic weight management.
	
	\subsection{NYSE-Only Universe}
	
	To limit concerns about historical NASDAQ volume coverage and exchange composition, I reconstruct the portfolios using only NYSE common stocks. The sample period, delisting treatment, factor data, and portfolio-generation procedure are unchanged. The selection ratio is 5\%, and each of the eight combinations of sampling and initial weighting contains 500 portfolios.
	
	\subsubsection{Buy and Hold}
	
	Table~\ref{tab:nyse-bh-cell} reports mean absolute pricing errors and significant-alpha shares for the NYSE-only BH portfolios.
	
	\FloatBarrier
	\begin{table}[H]
		\centering
		\onehalfspacing
		\caption{Factor-Model Performance on NYSE-Only Random Portfolios:
			Buy and Hold, 5\% Selection Ratio}
		\label{tab:nyse-bh-cell}
		\small
		\setlength{\tabcolsep}{5.0pt}
		
		\begin{tabular}{llrrrrrr}
			\toprule
			Sampling & Initial weight
			& CAPM & FF3 & Carhart & FF5 & FF6 & q5 \\
			\midrule
			
			\multicolumn{8}{l}{
				\textit{Panel A: Mean Absolute Pricing Error (bp)}
			} \\
			
			MEP  & EW
			& 33.4
			& 57.9
			& 50.9
			& 168.3
			& 150.3
			& \textbf{33.3} \\
			
			MEP  & VW
			& 95.1
			& 95.0
			& \textbf{86.4}
			& 256.4
			& 232.7
			& 143.5 \\
			
			MEP  & SQME
			& \textbf{49.1}
			& 76.3
			& 68.6
			& 212.4
			& 192.1
			& 66.4 \\
			
			MEP  & RAND
			& \textbf{37.3}
			& 61.4
			& 54.0
			& 169.2
			& 150.9
			& 38.0 \\
			
			\addlinespace
			
			UNIF & EW
			& 174.1
			& 63.9
			& \textbf{51.8}
			& 139.8
			& 111.4
			& 123.2 \\
			
			UNIF & VW
			& \textbf{72.1}
			& 93.8
			& 78.9
			& 180.0
			& 156.5
			& 77.5 \\
			
			UNIF & SQME
			& 95.1
			& 60.9
			& \textbf{51.3}
			& 146.7
			& 125.0
			& 82.0 \\
			
			UNIF & RAND
			& 175.2
			& 70.1
			& \textbf{59.0}
			& 138.6
			& 111.0
			& 126.0 \\
			
			\midrule
			
			\multicolumn{8}{l}{
				\textit{Panel B: Share of Significant Alphas (\%)}
			} \\
			
			MEP  & EW
			& \textbf{0.2}
			& 9.6
			& 4.4
			& 96.2
			& 91.8
			& 1.4 \\
			
			MEP  & VW
			& \textbf{6.0}
			& 32.2
			& 16.4
			& 100.0
			& 100.0
			& 83.6 \\
			
			MEP  & SQME
			& \textbf{2.0}
			& 29.6
			& 15.0
			& 100.0
			& 100.0
			& 13.6 \\
			
			MEP  & RAND
			& \textbf{0.2}
			& 10.4
			& 5.2
			& 90.6
			& 81.6
			& 1.2 \\
			
			\addlinespace
			
			UNIF & EW
			& 23.8
			& 5.8
			& \textbf{2.0}
			& 47.8
			& 28.6
			& 23.8 \\
			
			UNIF & VW
			& \textbf{2.8}
			& 6.8
			& 3.4
			& 44.8
			& 34.8
			& 4.4 \\
			
			UNIF & SQME
			& 11.2
			& 6.8
			& \textbf{3.2}
			& 60.0
			& 43.6
			& 11.8 \\
			
			UNIF & RAND
			& 20.6
			& 4.6
			& \textbf{2.2}
			& 40.4
			& 23.4
			& 21.4 \\
			
			\bottomrule
		\end{tabular}
		
		\vspace{0.4em}
		\begin{minipage}{0.95\textwidth}
			\footnotesize
			\textit{Notes:} Each design contains 500 BH portfolios. Panel A reports
			the cross-sectional mean of annualized absolute pricing errors in basis
			points. Panel B reports the percentage of portfolios with an alpha
			significant at the 5\% level using Newey--West standard errors.
			Boldface denotes the minimum in each row.
		\end{minipage}
	\end{table}
	\FloatBarrier
	
	The NYSE-only BH results again produce no construction-invariant ranking. Carhart has the smallest mean absolute pricing error in four cells, the CAPM in three, and q5 in one. q5 performs nearly identically to the CAPM in MEP--EW, but deteriorates sharply in MEP--VW.
	
	Among UNIF portfolios, Carhart performs best under EW, SQME, and RAND, while the CAPM performs best under VW. FF5 and FF6 perform poorly, especially in the MEP cells. Their relative strength in the full-exchange BH analysis therefore does not survive the NYSE-only restriction.
	
	\subsubsection{Constant Weight}
	
	Table~\ref{tab:nyse-cw-cell} reports the corresponding results under daily rebalancing.
	
	\FloatBarrier
	\begin{table}[H]
		\centering
		\onehalfspacing
		\caption{Factor-Model Performance on NYSE-Only Random Portfolios:
			Constant Weight, 5\% Selection Ratio}
		\label{tab:nyse-cw-cell}
		\small
		\setlength{\tabcolsep}{5.0pt}
		
		\begin{tabular}{llrrrrrr}
			\toprule
			Sampling & Initial weight
			& CAPM & FF3 & Carhart & FF5 & FF6 & q5 \\
			\midrule
			
			\multicolumn{8}{l}{
				\textit{Panel A: Mean Absolute Pricing Error (bp)}
			} \\
			
			MEP  & EW
			& 67.4
			& \textbf{36.0}
			& 57.5
			& 110.7
			& 44.8
			& 136.6 \\
			
			MEP  & VW
			& 60.9
			& 69.2
			& \textbf{20.4}
			& 207.0
			& 141.7
			& 44.5 \\
			
			MEP  & SQME
			& \textbf{23.7}
			& 47.8
			& 30.0
			& 157.4
			& 89.4
			& 47.1 \\
			
			MEP  & RAND
			& 66.9
			& \textbf{41.3}
			& 57.9
			& 111.9
			& 48.6
			& 134.9 \\
			
			\addlinespace
			
			UNIF & EW
			& 284.4
			& 61.4
			& 145.8
			& \textbf{52.3}
			& 73.1
			& 309.6 \\
			
			UNIF & VW
			& 87.4
			& 82.4
			& \textbf{79.0}
			& 132.1
			& 83.7
			& 128.4 \\
			
			UNIF & SQME
			& 155.6
			& 48.0
			& 92.8
			& 83.4
			& \textbf{45.5}
			& 210.2 \\
			
			UNIF & RAND
			& 285.1
			& 68.3
			& 147.5
			& \textbf{59.7}
			& 79.6
			& 309.3 \\
			
			\midrule
			
			\multicolumn{8}{l}{
				\textit{Panel B: Share of Significant Alphas (\%)}
			} \\
			
			MEP  & EW
			& 5.8
			& \textbf{0.6}
			& 7.0
			& 54.0
			& 2.2
			& 65.4 \\
			
			MEP  & VW
			& 0.4
			& 4.4
			& \textbf{0.0}
			& 100.0
			& 88.8
			& 0.8 \\
			
			MEP  & SQME
			& \textbf{0.0}
			& 4.4
			& 0.6
			& 98.2
			& 44.8
			& 2.0 \\
			
			MEP  & RAND
			& 6.6
			& \textbf{1.0}
			& 5.8
			& 43.2
			& 4.2
			& 50.8 \\
			
			\addlinespace
			
			UNIF & EW
			& 82.6
			& 2.4
			& 42.6
			& \textbf{2.0}
			& 7.2
			& 98.6 \\
			
			UNIF & VW
			& 5.2
			& 4.4
			& \textbf{4.0}
			& 18.2
			& 4.4
			& 17.0 \\
			
			UNIF & SQME
			& 38.4
			& \textbf{2.4}
			& 13.4
			& 12.8
			& 2.8
			& 77.8 \\
			
			UNIF & RAND
			& 75.2
			& \textbf{3.4}
			& 33.2
			& 3.6
			& 7.6
			& 94.0 \\
			
			\bottomrule
		\end{tabular}
		
		\vspace{0.4em}
		\begin{minipage}{0.95\textwidth}
			\footnotesize
			\textit{Notes:} Each design contains 500 CW portfolios. Panel A reports
			the cross-sectional mean of annualized absolute pricing errors in basis
			points. Panel B reports the percentage of portfolios with an alpha
			significant at the 5\% level using Newey--West standard errors.
			Boldface denotes the minimum in each row.
		\end{minipage}
	\end{table}
	\FloatBarrier
	
	Under CW, the row minima are distributed across FF3, Carhart, the CAPM, FF5, and FF6. FF3 performs best in MEP--EW and MEP--RAND, Carhart in both VW cells, FF5 in UNIF--EW and UNIF--RAND, and FF6 in UNIF--SQME. q5 does not rank first in any cell. It remains competitive in MEP--VW but leaves large errors in the other MEP designs and in most UNIF designs.
	
	The BH--CW comparison also changes by model. q5 favors BH in six of the eight cells but favors CW in MEP--VW and MEP--SQME. By contrast, FF3, FF5, and FF6 have smaller mean absolute errors under CW in all eight cells. Block-bootstrap inference shows a similar contrast: under BH, the mean-alpha null is rejected in all eight cells for FF5 and FF6 but in only two for q5. Under CW, q5 is rejected in six cells, FF5 in five, and FF3 in none.
	
	Restricting the universe to NYSE stocks therefore neither uniformly strengthens nor weakens the baseline results. q5 improves in some MEP BH designs, while FF5 and FF6 deteriorate. Under CW, rankings disperse across models, but q5 continues to leave large errors in the UNIF portfolios. The broader conclusion remains unchanged: factor-model performance depends on the interaction among stock selection, initial weighting, and post-formation weight management.
	
	\subsection{NYSE 20\% Market-Equity Breakpoint}
	
	The second robustness test retains NYSE, AMEX, and NASDAQ stocks but excludes firms below the twentieth percentile of the NYSE market-equity distribution at each formation date. This restriction preserves exchange coverage while limiting the influence of microcaps. I denote market-equity-proportional and uniform sampling after the filter by MEP20 and UNIF20.
	
	The sample period, delisting treatment, factor data, and portfolio-generation procedure are unchanged. The stock-selection ratio is 5\%, and each of the eight combinations of sampling and initial weighting contains 500 portfolios.
	
	\subsubsection{Buy and Hold}
	
	Table~\ref{tab:robust-nyse20-bh-cell} reports the BH results.
	
	\FloatBarrier
	\begin{table}[H]
		\centering
		\onehalfspacing
		\caption{NYSE 20\% Market-Equity Breakpoint:
			Buy and Hold, 5\% Selection Ratio}
		\label{tab:robust-nyse20-bh-cell}
		\small
		\setlength{\tabcolsep}{5.0pt}
		
		\begin{tabular}{llrrrrrr}
			\toprule
			Sampling & Initial weight
			& CAPM & FF3 & Carhart & FF5 & FF6 & q5 \\
			\midrule
			
			\multicolumn{8}{l}{
				\textit{Panel A: Mean Absolute Pricing Error (bp)}
			} \\
			
			MEP20  & EW
			& \textbf{29.2}
			& 35.5
			& 32.1
			& 40.2
			& 36.9
			& 43.0 \\
			
			MEP20  & VW
			& 34.5
			& 51.3
			& 56.8
			& 19.3
			& \textbf{17.2}
			& 145.2 \\
			
			MEP20  & SQME
			& \textbf{24.2}
			& 25.5
			& 25.1
			& 30.0
			& 25.7
			& 50.3 \\
			
			MEP20  & RAND
			& \textbf{35.4}
			& 39.9
			& 37.2
			& 44.4
			& 41.2
			& 47.4 \\
			
			\addlinespace
			
			UNIF20 & EW
			& 69.1
			& 82.0
			& 70.8
			& 50.9
			& \textbf{45.7}
			& 99.2 \\
			
			UNIF20 & VW
			& 69.4
			& 74.0
			& \textbf{65.9}
			& 80.4
			& 74.8
			& 77.0 \\
			
			UNIF20 & SQME
			& 50.8
			& 58.5
			& 52.8
			& 49.7
			& \textbf{46.0}
			& 79.2 \\
			
			UNIF20 & RAND
			& 72.0
			& 86.5
			& 74.0
			& 59.5
			& \textbf{52.9}
			& 100.3 \\
			
			\midrule
			
			\multicolumn{8}{l}{
				\textit{Panel B: Share of Significant Alphas (\%)}
			} \\
			
			MEP20  & EW
			& 4.4
			& 5.8
			& \textbf{4.2}
			& 11.4
			& 8.2
			& 14.2 \\
			
			MEP20  & VW
			& \textbf{0.0}
			& 7.2
			& 8.0
			& \textbf{0.0}
			& \textbf{0.0}
			& 99.6 \\
			
			MEP20  & SQME
			& \textbf{1.6}
			& 6.4
			& 6.4
			& 11.0
			& 5.4
			& 34.6 \\
			
			MEP20  & RAND
			& 4.2
			& 6.4
			& \textbf{3.6}
			& 9.8
			& 6.4
			& 13.6 \\
			
			\addlinespace
			
			UNIF20 & EW
			& \textbf{1.0}
			& 24.6
			& 17.4
			& 8.6
			& 5.4
			& 34.4 \\
			
			UNIF20 & VW
			& 4.2
			& 7.2
			& \textbf{3.6}
			& 8.2
			& 7.0
			& 6.6 \\
			
			UNIF20 & SQME
			& \textbf{3.6}
			& 11.4
			& 8.2
			& 7.6
			& 5.0
			& 26.2 \\
			
			UNIF20 & RAND
			& \textbf{1.4}
			& 23.2
			& 15.2
			& 8.6
			& 6.6
			& 28.0 \\
			
			\bottomrule
		\end{tabular}
		
		\vspace{0.4em}
		\begin{minipage}{0.95\textwidth}
			\footnotesize
			\textit{Notes:} Each design contains 500 BH portfolios. Panel A reports
			the cross-sectional mean of annualized absolute pricing errors in basis
			points. Panel B reports the percentage of portfolios with an alpha
			significant at the 5\% level using Newey--West standard errors.
			Boldface denotes the minimum in each row.
		\end{minipage}
	\end{table}
	\FloatBarrier
	
	The BH minima remain dispersed across models. The CAPM performs best in three MEP20 cells, FF6 in MEP20--VW and three UNIF20 cells, and Carhart in UNIF20--VW. Removing microcaps therefore does not produce a stable ranking.
	
	q5 does not rank first in any cell. It remains competitive in MEP20--EW and MEP20--RAND, but performs poorly in MEP20--VW and several UNIF20 designs. Relative to the baseline universe, its error falls in only two cells and rises in six. The microcap filter therefore does not uniformly improve BH performance.
	
	\subsubsection{Constant Weight}
	
	Table~\ref{tab:robust-nyse20-cw-cell} reports the CW results.
	
	\FloatBarrier
	\begin{table}[H]
		\centering
		\onehalfspacing
		\caption{NYSE 20\% Market-Equity Breakpoint:
			Constant Weight, 5\% Selection Ratio}
		\label{tab:robust-nyse20-cw-cell}
		\small
		\setlength{\tabcolsep}{5.0pt}
		
		\begin{tabular}{llrrrrrr}
			\toprule
			Sampling & Initial weight
			& CAPM & FF3 & Carhart & FF5 & FF6 & q5 \\
			\midrule
			
			\multicolumn{8}{l}{
				\textit{Panel A: Mean Absolute Pricing Error (bp)}
			} \\
			
			MEP20  & EW
			& 44.1
			& \textbf{30.3}
			& 96.2
			& 36.8
			& 100.5
			& 179.8 \\
			
			MEP20  & VW
			& \textbf{18.5}
			& 69.9
			& 146.0
			& 41.0
			& 103.9
			& 30.0 \\
			
			MEP20  & SQME
			& \textbf{22.5}
			& 39.5
			& 120.4
			& 32.2
			& 95.7
			& 80.1 \\
			
			MEP20  & RAND
			& 47.3
			& \textbf{35.7}
			& 97.3
			& 42.1
			& 102.0
			& 180.4 \\
			
			\addlinespace
			
			UNIF20 & EW
			& 124.2
			& \textbf{49.6}
			& 82.8
			& 50.8
			& 125.3
			& 271.9 \\
			
			UNIF20 & VW
			& 75.7
			& \textbf{73.8}
			& 110.2
			& 78.8
			& 111.2
			& 149.7 \\
			
			UNIF20 & SQME
			& 81.3
			& \textbf{45.0}
			& 88.7
			& 50.8
			& 114.5
			& 230.9 \\
			
			UNIF20 & RAND
			& 122.9
			& \textbf{56.1}
			& 84.1
			& 56.6
			& 124.8
			& 270.9 \\
			
			\midrule
			
			\multicolumn{8}{l}{
				\textit{Panel B: Share of Significant Alphas (\%)}
			} \\
			
			MEP20  & EW
			& 7.4
			& \textbf{1.6}
			& 51.0
			& 5.4
			& 57.4
			& 98.8 \\
			
			MEP20  & VW
			& \textbf{0.0}
			& 13.6
			& 99.2
			& 1.2
			& 69.8
			& \textbf{0.0} \\
			
			MEP20  & SQME
			& \textbf{0.0}
			& 8.6
			& 98.0
			& 4.8
			& 81.6
			& 48.6 \\
			
			MEP20  & RAND
			& 6.4
			& \textbf{1.6}
			& 43.0
			& 4.4
			& 47.0
			& 96.0 \\
			
			\addlinespace
			
			UNIF20 & EW
			& 11.2
			& \textbf{2.4}
			& 18.6
			& 4.8
			& 51.4
			& 99.6 \\
			
			UNIF20 & VW
			& 5.4
			& \textbf{5.2}
			& 16.6
			& 6.6
			& 17.0
			& 32.0 \\
			
			UNIF20 & SQME
			& 10.0
			& \textbf{2.2}
			& 24.8
			& 5.8
			& 44.8
			& 96.8 \\
			
			UNIF20 & RAND
			& 10.0
			& \textbf{3.2}
			& 16.6
			& 5.2
			& 42.4
			& 97.2 \\
			
			\bottomrule
		\end{tabular}
		
		\vspace{0.4em}
		\begin{minipage}{0.95\textwidth}
			\footnotesize
			\textit{Notes:} Each design contains 500 CW portfolios. Panel A reports
			the cross-sectional mean of annualized absolute pricing errors in basis
			points. Panel B reports the percentage of portfolios with an alpha
			significant at the 5\% level using Newey--West standard errors.
			Boldface denotes the minimum in each row.
		\end{minipage}
	\end{table}
	\FloatBarrier
	
	FF3 is the most stable CW model. It produces the smallest mean absolute pricing error in six of the eight cells. The CAPM performs best in MEP20--VW and MEP20--SQME.
	
	q5 remains competitive in MEP20--VW, with a 30.0 bp mean absolute error and no significant alphas, but does not rank first in any cell. Its errors remain especially large in MEP20--EW, MEP20--RAND, and several UNIF20 designs.
	
	The microcap filter generally reduces CW errors. Relative to the baseline universe, q5 improves in seven cells and FF6 in all eight. Even after the filter, however, q5 continues to leave large positive pricing errors in several UNIF20 portfolios. Microcaps amplify the errors, but do not fully explain the model differences.
	
	\subsubsection{Weight Management and Portfolio Breadth}
	
	The BH--CW comparison remains model- and design-dependent. q5 favors BH in seven cells and CW only in MEP20--VW, where CW outperforms BH for all 500 matched portfolios. FF3 favors CW in six cells, whereas Carhart and FF6 favor BH in all eight. No weight-management rule is uniformly advantageous.
	
	Increasing portfolio breadth does not remove the construction dependence. Under BH, FF6 becomes stronger at high selection ratios, while q5 leaves larger errors than FF3 in every design from 5\% through 90\%. Under CW, FF3 becomes even more dominant and ranks first in all eight cells at the 90\% selection ratio. q5 exceeds FF3 in every cell once the selection ratio reaches 25\%.
	
	Block-bootstrap inference shows the same asymmetry. Under BH, the q5 mean-alpha null is rejected in five cells; under CW, it is rejected in seven. FF6 is never rejected under BH but is rejected in all eight CW cells. FF3 is rejected in three BH cells and in none of the CW cells.
	
	\subsubsection{Summary}
	
	Applying the NYSE twentieth-percentile breakpoint reduces many CW pricing errors, confirming that microcaps affect their magnitude. The improvement is neither uniform across models nor present in every BH design.
	
	After the filter, BH rankings remain dispersed across the CAPM, Carhart, and FF6. Under CW, FF3 remains the most stable model. q5 does not rank first in any BH or CW cell and continues to leave large positive errors in several daily rebalancing designs.
	
	The baseline findings therefore cannot be attributed solely to extreme microcaps. Their removal changes the scale of some errors, but model performance remains sensitive to sampling, initial weighting, portfolio breadth, and weight management.
	
	\subsection{Portfolio Breadth and Pricing Errors}
	
	The baseline analysis fixes the stock-selection ratio at 5\%. This section instead
	holds the ALL--UNIF--VW design fixed and traces pricing errors as portfolio breadth
	increases. Selection ratios range from 0.5\% to 100\%. For ratios below 100\%,
	each weight-management rule contains 500 portfolios. At 100\%, every eligible
	stock is included, leaving a single ALL--VW portfolio; this observation is therefore
	a boundary point rather than part of the Monte Carlo cross-section.

	\subsubsection{Buy--and--Hold}
	
	Table~\ref{tab:ratio-unif-vw-bh} reports the BH results.
	
	\FloatBarrier
	\begin{table}[H]
		\centering
		\onehalfspacing
		\caption{Selection-Ratio Analysis for ALL--UNIF--VW:
			Buy--and--Hold}
		\label{tab:ratio-unif-vw-bh}
		\small
		\setlength{\tabcolsep}{5.5pt}
		
		\begin{tabular}{rrrrrrr}
			\toprule
			Selection ratio
			& CAPM & FF3 & Carhart & FF5 & FF6 & q5 \\
			\midrule
			
			\multicolumn{7}{l}{
				\textit{Panel A: Mean Absolute Pricing Error (bp)}
			} \\
			
			0.5\%
			& 156.1 & 162.5 & \textbf{145.6}
			& 164.5 & 158.2 & 175.6 \\
			
			1\%
			& 124.2 & 132.0 & \textbf{115.6}
			& 134.1 & 126.5 & 141.9 \\
			
			5\%
			& 67.6 & 72.1 & \textbf{62.7}
			& 76.3 & 70.5 & 73.3 \\
			
			10\%
			& 50.7 & 54.1 & \textbf{48.1}
			& 57.4 & 53.9 & 54.8 \\
			
			25\%
			& 31.3 & 33.7 & \textbf{30.2}
			& 37.9 & 35.5 & 35.4 \\
			
			50\%
			& 18.7 & 20.4 & \textbf{17.8}
			& 27.1 & 23.9 & 24.9 \\
			
			75\%
			& 11.5 & 12.7 & \textbf{10.7}
			& 22.5 & 18.9 & 20.7 \\
			
			90\%
			& 7.0 & 8.3 & \textbf{6.6}
			& 22.0 & 18.2 & 20.4 \\
			
			100\%
			& 3.5 & 5.9 & \textbf{3.4}
			& 21.6 & 17.7 & 20.1 \\
			
			\midrule
			
			\multicolumn{7}{l}{
				\textit{Panel B: Share of Significant Alphas (\%)}
			} \\
			
			0.5\%
			& 4.6 & 5.4 & 4.2
			& 6.2 & \textbf{4.0} & 8.2 \\
			
			1\%
			& 4.6 & 6.4 & \textbf{3.8}
			& 7.0 & 5.8 & 8.6 \\
			
			5\%
			& 5.0 & 6.4 & \textbf{3.2}
			& 8.0 & 5.6 & 4.6 \\
			
			10\%
			& 4.2 & 5.2 & \textbf{2.4}
			& 8.6 & 6.6 & 5.0 \\
			
			25\%
			& 5.4 & 6.2 & \textbf{3.8}
			& 10.0 & 6.6 & 7.8 \\
			
			50\%
			& \textbf{3.6} & 6.0 & \textbf{3.6}
			& 16.0 & 10.2 & 12.0 \\
			
			75\%
			& 2.0 & 4.6 & \textbf{0.8}
			& 29.0 & 18.2 & 20.2 \\
			
			90\%
			& 0.2 & 1.0 & \textbf{0.0}
			& 52.6 & 32.2 & 40.0 \\
			
			100\%
			& \textbf{0.0} & \textbf{0.0} & \textbf{0.0}
			& 100.0 & 100.0 & 100.0 \\
			
			\bottomrule
		\end{tabular}
		
		\vspace{0.4em}
		\begin{minipage}{0.95\textwidth}
			\footnotesize
			\textit{Notes:} Each selection ratio below 100\% contains 500
			ALL--UNIF--VW Buy--and--Hold portfolios. At 100\%, only one
			ALL--VW portfolio exists. Panel A reports the cross-sectional mean of
			annualized absolute pricing errors. Panel B reports the percentage of
			portfolios with an alpha significant at the 5\% level using Newey--West
			standard errors. Values of 0\% and 100\% in the final row indicate
			insignificance or significance for the single portfolio. Boldface denotes
			the minimum in each row.
		\end{minipage}
	\end{table}
	\FloatBarrier
	
	Mean absolute errors decline with breadth for every model, but at different rates.
	Carhart ranks first throughout and falls to 3.4 bp at the 100\% boundary. FF5,
	FF6, and q5 retain errors of 21.6, 17.7, and 20.1 bp, respectively.
	
	The divergence also becomes more systematic across portfolios. The share for which
	FF3 has a smaller absolute error than q5 rises from 62.0\% at a 50\% selection
	ratio to 73.8\% at 75\% and 94.2\% at 90\%. Over the same range, q5's
	significant-alpha share rises from 12.0\% to 40.0\%. Diversification therefore
	reduces stock-specific noise without eliminating common model-specific pricing errors.

	\subsubsection{Constant--Weight}
	
	Table~\ref{tab:ratio-unif-vw-cw} reports the corresponding daily
	constant-weight results.
	
	\FloatBarrier
	\begin{table}[H]
		\centering
		\onehalfspacing
		\caption{Selection-Ratio Analysis for ALL--UNIF--VW:
			Constant--Weight}
		\label{tab:ratio-unif-vw-cw}
		\small
		\setlength{\tabcolsep}{5.5pt}
		
		\begin{tabular}{rrrrrrr}
			\toprule
			Selection ratio
			& CAPM & FF3 & Carhart & FF5 & FF6 & q5 \\
			\midrule
			
			\multicolumn{7}{l}{
				\textit{Panel A: Mean Absolute Pricing Error (bp)}
			} \\
			
			0.5\%
			& 170.7 & \textbf{163.2} & 171.3
			& 164.4 & 187.2 & 258.1 \\
			
			1\%
			& 136.4 & \textbf{131.5} & 144.9
			& 133.0 & 158.5 & 222.5 \\
			
			5\%
			& 78.3 & \textbf{72.7} & 114.0
			& 77.6 & 115.9 & 158.2 \\
			
			10\%
			& 61.2 & \textbf{55.6} & 113.8
			& 59.9 & 110.6 & 142.5 \\
			
			25\%
			& 45.1 & \textbf{38.7} & 114.1
			& 46.0 & 109.8 & 131.5 \\
			
			50\%
			& 36.0 & \textbf{27.9} & 113.2
			& 39.2 & 109.5 & 126.9 \\
			
			75\%
			& 34.0 & \textbf{23.8} & 113.6
			& 38.2 & 109.7 & 124.8 \\
			
			90\%
			& 34.1 & \textbf{23.5} & 113.6
			& 38.1 & 109.5 & 124.4 \\
			
			100\%
			& 34.2 & \textbf{23.6} & 114.0
			& 38.5 & 110.0 & 124.6 \\
			
			\midrule
			
			\multicolumn{7}{l}{
				\textit{Panel B: Share of Significant Alphas (\%)}
			} \\
			
			0.5\%
			& \textbf{4.8} & 6.0 & \textbf{4.8}
			& 5.4 & 7.8 & 18.6 \\
			
			1\%
			& 7.0 & \textbf{5.6} & 7.2
			& 6.8 & 10.4 & 21.6 \\
			
			5\%
			& 6.0 & \textbf{5.0} & 18.8
			& 6.6 & 19.2 & 37.0 \\
			
			10\%
			& 7.6 & \textbf{5.4} & 34.2
			& 7.2 & 35.2 & 49.4 \\
			
			25\%
			& 8.6 & \textbf{6.2} & 73.6
			& 9.4 & 68.8 & 78.2 \\
			
			50\%
			& 7.0 & \textbf{3.8} & 98.6
			& 8.8 & 97.2 & 97.8 \\
			
			75\%
			& 2.6 & \textbf{0.8} & 100.0
			& 4.4 & 100.0 & 100.0 \\
			
			90\%
			& 0.4 & \textbf{0.0} & 100.0
			& 1.4 & 100.0 & 100.0 \\
			
			100\%
			& \textbf{0.0} & \textbf{0.0} & 100.0
			& \textbf{0.0} & 100.0 & 100.0 \\
			
			\bottomrule
		\end{tabular}
		
		\vspace{0.4em}
		\begin{minipage}{0.95\textwidth}
			\footnotesize
			\textit{Notes:} Each selection ratio below 100\% contains 500
			ALL--UNIF--VW Constant--Weight portfolios. At 100\%, only one
			ALL--VW portfolio exists. Panel A reports the cross-sectional mean of
			annualized absolute pricing errors. Panel B reports the percentage of
			portfolios with an alpha significant at the 5\% level using Newey--West
			standard errors. Values of 0\% and 100\% in the final row indicate
			insignificance or significance for the single portfolio. Boldface denotes
			the minimum in each row.
		\end{minipage}
	\end{table}
	\FloatBarrier
	
	FF3 has the smallest mean absolute error at every CW selection ratio. As breadth
	increases, CAPM and FF5 converge to roughly 34--39 bp, whereas Carhart and FF6
	plateau near 110 bp and q5 near 125 bp.
	
	The significant-alpha shares show the same separation. q5 reaches 97.8\% at 50\%
	and 100\% at ratios of 75\% or more. Carhart and FF6 also reach 100\% above
	75\%, while FF3 has no significant alphas at 90\%. FF3 has a smaller absolute
	error than q5 for every matched portfolio at selection ratios of 50\% or more.

	\subsubsection{Summary}
	
	Greater breadth reduces cross-sectional sampling noise, but the convergence path
	depends sharply on weight management. Under BH, all models improve and Carhart
	remains best, although FF5, FF6, and q5 retain nontrivial errors at the 100\%
	boundary. Under CW, only FF3 converges to a low error level; Carhart, FF6, and
	q5 retain large positive pricing errors.
	
	The persistence of these differences at 100\% shows that the results are not driven
	solely by random stock selection or a few extreme portfolios. Because the exercise
	fixes the ALL--UNIF--VW design, however, it should not be interpreted as a universal
	ranking across all sampling and weighting rules. Its implication is narrower:
	portfolio breadth and weight management jointly determine how factor-model pricing
	errors converge.
	
	\subsection{Rebalancing-Frequency Dependence}
	
	To determine whether the BH--CW contrast is specific to two endpoint rules, I vary
	rebalancing frequency across annual, quarterly, monthly, weekly, and daily schedules.
	The stock-selection ratio is fixed at 5\%, and the same stocks and initial target
	weights are used at every frequency. Differences therefore arise from post-formation
	weight paths rather than static portfolio composition.
	
	Annual rebalancing corresponds to BH. Quarterly and monthly portfolios restore
	target weights on the first trading day of each quarter or month. Weekly portfolios
	rebalance before Wednesday's return, or on the next trading day when Wednesday is
	a holiday. Daily rebalancing corresponds to CW. After delisting, a stock's target
	weight is assigned to a cash account earning the risk-free rate. The analysis tracks
	the same 500 portfolios across all five frequencies for each of eight combinations
	of MEP or UNIF sampling and four initial-weighting rules.
	
	The main text reports UNIF--EW and UNIF--VW as representative cases; the other six
	designs are used to assess generality and exceptions.

	\subsubsection{Mean Absolute Pricing Errors}
	
	Table~\ref{tab:rebalance-unif-abs} reports mean absolute pricing errors for the two
	representative designs.
	
	\FloatBarrier
	\begin{table}[H]
		\centering
		\onehalfspacing
		\caption{Mean Absolute Pricing Errors by Rebalancing Frequency}
		\label{tab:rebalance-unif-abs}
		\small
		\setlength{\tabcolsep}{6pt}
		
		\begin{tabular}{lrrrrrr}
			\toprule
			Rebalancing frequency
			& CAPM & FF3 & Carhart & FF5 & FF6 & q5 \\
			\midrule
			
			\multicolumn{7}{l}{
				\textit{Panel A: UNIF--EW}
			} \\
			
			ANNUAL
			& 80.0
			& 107.1
			& 102.2
			& 61.5
			& \textbf{59.3}
			& 94.8 \\
			
			QUARTERLY
			& 52.1
			& 152.3
			& 62.3
			& 91.7
			& \textbf{40.4}
			& 143.4 \\
			
			MONTHLY
			& 55.0
			& 149.7
			& 50.6
			& 84.6
			& \textbf{37.7}
			& 163.7 \\
			
			WEEKLY
			& 169.0
			& \textbf{44.8}
			& 80.8
			& 50.0
			& 125.3
			& 291.3 \\
			
			DAILY
			& 368.3
			& \textbf{169.1}
			& 276.4
			& 241.8
			& 324.2
			& 490.3 \\
			
			\midrule
			
			\multicolumn{7}{l}{
				\textit{Panel B: UNIF--VW}
			} \\
			
			ANNUAL
			& 67.6
			& 72.1
			& \textbf{62.7}
			& 76.3
			& 70.5
			& 73.3 \\
			
			QUARTERLY
			& \textbf{68.6}
			& 76.2
			& 73.3
			& 76.1
			& 75.1
			& 97.1 \\
			
			MONTHLY
			& \textbf{69.8}
			& 79.2
			& 74.3
			& 76.0
			& 77.3
			& 102.6 \\
			
			WEEKLY
			& \textbf{71.1}
			& 71.2
			& 97.9
			& 73.0
			& 99.9
			& 137.7 \\
			
			DAILY
			& 78.3
			& \textbf{72.7}
			& 114.0
			& 77.6
			& 115.9
			& 158.2 \\
			
			\bottomrule
		\end{tabular}
		
		\vspace{0.4em}
		\begin{minipage}{0.95\textwidth}
			\footnotesize
			\textit{Notes:} Each design and frequency uses the same 500 portfolios.
			Entries are cross-sectional means of annualized absolute pricing errors in
			basis points. Boldface denotes the minimum in each row.
		\end{minipage}
	\end{table}
	\FloatBarrier
	
	In UNIF--EW, FF6 ranks first from annual through monthly rebalancing, but FF3 ranks
	first at weekly and daily frequencies. FF3's daily error nevertheless exceeds its
	annual error, so the rank improvement partly reflects larger deterioration among
	competitors. In UNIF--VW, the minimum shifts from Carhart under annual rebalancing
	to CAPM from quarterly through weekly and to FF3 under daily rebalancing.
	
	Both designs show a reversal in the contribution of momentum. Carhart outperforms
	FF3 and FF6 outperforms FF5 under annual rebalancing, whereas the ordering reverses
	at weekly and daily frequencies.

	\subsubsection{Significant-Alpha Shares}
	
	Table~\ref{tab:rebalance-unif-sig} reports the corresponding significant-alpha
	shares.
	
	\FloatBarrier
	\begin{table}[H]
		\centering
		\onehalfspacing
		\caption{Significant-Alpha Shares by Rebalancing Frequency}
		\label{tab:rebalance-unif-sig}
		\small
		\setlength{\tabcolsep}{6pt}
		
		\begin{tabular}{lrrrrrr}
			\toprule
			Rebalancing frequency
			& CAPM & FF3 & Carhart & FF5 & FF6 & q5 \\
			\midrule
			
			\multicolumn{7}{l}{
				\textit{Panel A: UNIF--EW}
			} \\
			
			ANNUAL
			& \textbf{0.0}
			& 47.6
			& 44.0
			& 14.6
			& 11.8
			& 29.0 \\
			
			QUARTERLY
			& \textbf{0.0}
			& 71.4
			& 6.6
			& 24.8
			& 1.4
			& 54.6 \\
			
			MONTHLY
			& \textbf{0.0}
			& 64.0
			& 3.0
			& 18.4
			& 0.8
			& 68.6 \\
			
			WEEKLY
			& 7.6
			& \textbf{1.6}
			& 14.2
			& 4.0
			& 55.2
			& 99.6 \\
			
			DAILY
			& 99.8
			& \textbf{77.8}
			& 99.8
			& 98.6
			& 100.0
			& 100.0 \\
			
			\midrule
			
			\multicolumn{7}{l}{
				\textit{Panel B: UNIF--VW}
			} \\
			
			ANNUAL
			& 5.0
			& 6.4
			& \textbf{3.2}
			& 8.0
			& 5.6
			& 4.6 \\
			
			QUARTERLY
			& \textbf{3.4}
			& 5.0
			& 4.2
			& 6.4
			& 5.2
			& 12.2 \\
			
			MONTHLY
			& \textbf{3.6}
			& 6.2
			& 4.2
			& 6.2
			& 5.4
			& 15.0 \\
			
			WEEKLY
			& 5.0
			& \textbf{3.0}
			& 11.2
			& 4.2
			& 12.6
			& 28.4 \\
			
			DAILY
			& 6.0
			& \textbf{5.0}
			& 18.8
			& 6.6
			& 19.2
			& 37.0 \\
			
			\bottomrule
		\end{tabular}
		
		\vspace{0.4em}
		\begin{minipage}{0.95\textwidth}
			\footnotesize
			\textit{Notes:} Entries are percentages of portfolios with alphas significant
			at the 5\% level using Newey--West standard errors. Each design and
			frequency uses the same 500 portfolios. Boldface denotes the minimum in
			each row.
		\end{minipage}
	\end{table}
	\FloatBarrier
	
	The rejection rates exhibit the same rank shifts. In UNIF--EW, CAPM records no
	significant alphas at low frequencies, whereas FF3 ranks first under weekly
	rebalancing. Under daily rebalancing, even FF3 reaches 77.8\%, and the other models
	are rejected for nearly every portfolio.
	
	UNIF--VW has lower rejection rates, but q5 and the momentum models deteriorate as
	rebalancing becomes more frequent. q5 rises from 4.6\% under annual rebalancing to
	37.0\% under daily rebalancing; Carhart and FF6 reach 18.8\% and 19.2\%.
	FF3 retains the lowest rejection rate at weekly and daily frequencies.

	\subsubsection{Full Design and Exceptions}
	
	The rank shifts extend to the full set of eight designs. Under annual rebalancing,
	FF6 minimizes mean absolute errors in four designs and CAPM in three. The minima
	are more dispersed at quarterly and monthly frequencies; FF5 leads in three weekly
	designs, and FF3 in six daily designs.
	
	FF3's daily rank does not imply uniform improvement: its daily error is below its
	annual error in only three designs. By contrast, Carhart is worse under daily
	rebalancing in all eight designs, and FF6 is worse under both weekly and daily
	rebalancing in all eight.
	
	For q5, mean absolute errors increase monotonically with frequency in seven designs,
	and significant-alpha shares do so in six. The main exception is MEP--VW, where
	q5's error falls from 110.3 bp annually to 9.9 bp weekly and its rejection rate is
	zero from monthly rebalancing onward.
	
	Mean $R^{2}$ declines relative to annual rebalancing in all 192
	model--design--frequency comparisons. Yet FF3 and FF5 show lower pricing errors in
	many weekly designs, so changes in pricing errors cannot be explained by time-series
	fit alone.

	\subsubsection{Summary}
	
	Rebalancing frequency changes both pricing errors and model rankings for portfolios
	with identical stocks and initial weights. FF6 is relatively strong at low frequencies,
	FF5 leads most often under weekly rebalancing, and FF3 under daily rebalancing.
	q5 generally deteriorates with frequency, although MEP--VW follows the opposite
	path.
	
	The BH--CW contrast is therefore not confined to two endpoints. Intermediate
	rebalancing frequencies interact with sampling and initial weighting to generate
	distinct dynamic return processes. These results measure the sensitivity of
	factor-model evaluation to weight management; they are not transaction-cost-adjusted
	strategy returns or tests of maximum-Sharpe dominance.
	
	\section{Discussion}
	\label{sec:discuss}
	
	\subsection{Construction Dependence and Evaluation Criteria}
	
	The central result is that incomplete factor models do not admit a construction-invariant ranking of pricing performance. For static CW portfolios, the integrated squared pricing error under an asset-generating distribution $\mathcal{D}$ is
	\begin{equation}
		\mathcal{L}_{\mathcal{D}}^{CW}(M)
		=
		\boldsymbol{\alpha}_{M}^{\prime}
		Q_{\mathcal{D}}
		\boldsymbol{\alpha}_{M}.
	\end{equation}
	The matrix $Q_{\mathcal{D}}$ reflects stock-selection probabilities, portfolio breadth, and initial weighting. It determines which directions of the underlying pricing-error vector receive the greatest weight. If the model were exact, this choice would be irrelevant. For incomplete models that fail in different directions, the evaluation distribution becomes part of the observed ranking.
	
	BH and intermediate rebalancing frequencies require a dynamic extension. Their weights evolve with realized returns, so the weight-management rule generates a distinct tradable return process rather than merely reweighting a fixed pricing-error vector. Construction dependence therefore has two dimensions: static portfolio formation and dynamic weight evolution.
	
	This distinction also separates factor spanning from random-portfolio evaluation. Spanning tests assess the mean--variance opportunity set offered by tradable factors, whereas the random-portfolio tests measure the magnitude and stability of portfolio-level pricing errors under a prespecified asset-generating distribution \citep{BS17,FF18}. q5 can therefore offer the broadest sample opportunity set and still leave relatively large pricing errors on random portfolios. The two results concern different objective functions.
	
	The paper does not propose any particular $Q_{\mathcal{D}}$ as a uniquely correct economic loss function. Instead, it varies sampling, initial weighting, portfolio breadth, investment universe, and rebalancing to document how sensitive model evaluation is to the chosen environment.
	
	\subsection{Dynamic Construction and Factor Heterogeneity}
	
	Construction dependence is not specific to a single model. FF5 and FF6 perform relatively well under BH and low-frequency rebalancing, whereas FF3 is strongest and most stable under CW and high-frequency rebalancing. Carhart and FF6 lose much of their low-frequency advantage as rebalancing becomes more frequent. External fit is therefore not a monotonic function of the number of factors.
	
	q5 makes this contrast especially clear. It delivers the highest maximum Sharpe ratio in the spanning analysis but leaves systematic positive pricing errors in many CW designs. The persistence of these errors across sampling rules, initial weights, and portfolio breadth suggests that they are not driven solely by a few extreme portfolios or by firm-specific noise.
	
	The expected-growth factor illustrates the difference between opportunity-set expansion and external pricing fit. EG contributes substantially to q5's maximum Sharpe ratio in \cref{sec:span}, yet it increases pricing errors on some Fama--French characteristic sorts and high-frequency random portfolios. This does not imply that EG is economically redundant. It shows that expanding the mean--variance opportunity set and reducing pricing errors on external portfolios are distinct properties.
	
	The mechanism behind the positive q5 alphas is not identified here. One possibility is that broad long-only baskets load systematically on ROE and EG and that repeated target-weight restoration sustains those exposures. Decomposing portfolio-level pricing errors into constituent betas and endogenous weight paths is a natural direction for future work.
	
	\subsection{Implications for Factor-Model Evaluation}
	
	The evidence has three implications.
	
	First, empirical comparisons should report both direct factor spanning and pricing errors on external test assets. The former evaluates relative investment opportunities; the latter evaluates external pricing fit. Neither subsumes the other.
	
	Second, test-asset construction should be treated as part of the research design rather than as incidental preprocessing. Stock selection, initial weighting, portfolio breadth, holding rules, and rebalancing frequency all affect the return processes being priced. BH and CW, in particular, need not be interpreted as alternative weightings of the same static object.
	
	Third, high time-series $R^{2}$, a larger maximum Sharpe ratio, and small mean-return pricing errors are separate properties. A factor may explain more return variation or expand the sample opportunity set without reducing intercept errors on external portfolios. Covariance fit, investment-opportunity contribution, and mean-return fit should therefore be evaluated separately.
	
	\subsection{Limitations and Future Research}
	
	The spanning results rely on in-sample maximum squared Sharpe ratios. These statistics may be inflated by estimation error and data mining. They should therefore be interpreted as descriptions of the sample opportunity set, not as attainable out-of-sample investment performance.
	
	Daily and weekly rebalancing are also experimental rules that abstract from transaction costs and market impact. Their purpose is not to propose implementable strategies, but to reveal how the same stocks and initial weights can generate different test assets under different weight-management rules.
	
	The analysis is limited to long-only portfolios of U.S. common stocks. Whether the same construction dependence appears in international markets, other asset classes, short-enabled portfolios, or settings with realistic trading frictions remains open.
	
	Finally, mean absolute error, root-mean-square error, tail errors, and significant-alpha shares represent only a subset of possible loss functions. The economically relevant asset-generating distribution and loss function depend on the purpose of the SDF approximation. The contribution of this paper is not to identify a universal evaluation measure. It is to show systematically that the performance of incomplete factor models is conditional on both static test-asset construction and dynamic weight management.
	
	\section{Conclusion}
	\label{sec:conclusion}
	
	This paper evaluates major factor models on characteristic-unsorted random portfolios drawn from the CRSP investable universe. The design varies stock selection, initial weights, portfolio breadth, and weight management, while separately examining home-field bias on conventional characteristic-sorted assets and direct factor spanning among the candidate models.
	
	The results do not yield a construction-invariant ranking. Models perform relatively better on test assets close to their own research lineages, and direct spanning indicates that q5 provides the broadest sample mean--variance opportunity set. On random portfolios, however, FF5 and FF6 are relatively strong under BH and low-frequency rebalancing, whereas FF3 is the most stable model under high-frequency rebalancing. q5 leaves systematic positive pricing errors in many CW and high-frequency designs, and these errors do not disappear as portfolios become broader.
	
	These findings do not imply universal model superiority. Factor spanning and maximum Sharpe ratios evaluate the investment opportunities generated by tradable factors, whereas portfolio-level alphas measure the return-unit errors left by an SDF approximation under a prespecified asset-generating distribution and weight-management rule. q5 can therefore dominate on the former criterion while leaving larger errors on external portfolios. The expected-growth factor illustrates the same distinction: it substantially expands the sample opportunity set but increases pricing errors for some test-asset families.
	
	Test-asset construction should therefore be treated as part of the evaluation design rather than as a neutral preprocessing choice. Stock selection and initial weighting emphasize different directions of misspecification, while holding and rebalancing rules generate distinct tradable return processes. Evaluating incomplete factor models requires reporting both factor spanning and the level, direction, and stability of pricing errors across alternative portfolio-generation and weight-management rules.
	
	In short, the question ``which model?'' cannot be separated from ``which portfolios?'' or from how those portfolios are managed.
	
	\subsection*{Funding}
	This research did not receive any specific grant from funding agencies in the public, commercial, or not-for-profit sectors.
	
	\subsection*{Declaration of AI usage in manuscript preparation}
	During the preparation of this manuscript, the author used ChatGPT (OpenAI) and Claude (Anthropic) for language refinement and structural clarity.
	All outputs were reviewed and edited by the author, who takes full responsibility for the content.
	
	\subsection*{Declaration of interest}
	The author declares no competing interests.
	
	\newpage
	\onehalfspacing

	\newpage
	\begin{appendices}
		
		\section{NYSE Market-Capitalization Breakpoint within the NYSE-Only Universe}
		\label{sec:app-nyse20}
		
		This appendix repeats the NYSE-only analysis after excluding stocks below the
		twentieth percentile of the NYSE market-equity distribution at each formation
		date. The sample period, delisting treatment, factor data, and portfolio-generation
		procedure are unchanged. Market-equity-proportional and uniform sampling after
		the filter are denoted MEP20 and UNIF20. Each design selects 5\% of the eligible
		universe and contains 500 portfolios for each of the four initial-weighting rules.
		
		\subsection{Buy--and--Hold}
		
		Table~\ref{tab:app-nyse20-bh} reports mean absolute pricing errors and
		significant-alpha shares for the filtered NYSE-only BH portfolios.
		
		\FloatBarrier
		\begin{table}[H]
			\centering
			\onehalfspacing
			\caption{NYSE Market-Capitalization Breakpoint within the NYSE-Only Universe:
				Buy--and--Hold}
			\label{tab:app-nyse20-bh}
			\small
			\setlength{\tabcolsep}{5.0pt}
			
			\begin{tabular}{llrrrrrr}
				\toprule
				Sampling & Initial weight
				& CAPM & FF3 & Carhart & FF5 & FF6 & q5 \\
				\midrule
				
				\multicolumn{8}{l}{
					\textit{Panel A: Mean Absolute Pricing Error (bp)}
				} \\
				
				MEP20  & EW
				& \textbf{35.4}
				& 57.1
				& 50.7
				& 167.4
				& 149.6
				& 36.5 \\
				
				MEP20  & VW
				& 97.7
				& 95.9
				& \textbf{86.9}
				& 260.0
				& 235.9
				& 150.2 \\
				
				MEP20  & SQME
				& \textbf{52.2}
				& 76.4
				& 68.3
				& 215.0
				& 194.3
				& 73.0 \\
				
				MEP20  & RAND
				& \textbf{41.5}
				& 60.6
				& 53.6
				& 167.9
				& 149.8
				& 42.6 \\
				
				\addlinespace
				
				UNIF20 & EW
				& 154.0
				& 53.9
				& \textbf{44.3}
				& 130.6
				& 105.3
				& 119.0 \\
				
				UNIF20 & VW
				& \textbf{68.0}
				& 84.2
				& 74.1
				& 178.0
				& 158.0
				& 75.6 \\
				
				UNIF20 & SQME
				& 87.9
				& 55.8
				& \textbf{47.7}
				& 144.5
				& 124.6
				& 76.9 \\
				
				UNIF20 & RAND
				& 153.2
				& 60.2
				& \textbf{50.6}
				& 132.4
				& 108.0
				& 118.6 \\
				
				\midrule
				
				\multicolumn{8}{l}{
					\textit{Panel B: Share of Significant Alphas (\%)}
				} \\
				
				MEP20  & EW
				& \textbf{0.8}
				& 9.8
				& 4.4
				& 92.4
				& 85.4
				& 1.4 \\
				
				MEP20  & VW
				& \textbf{6.4}
				& 24.4
				& 11.6
				& 100.0
				& 100.0
				& 81.8 \\
				
				MEP20  & SQME
				& \textbf{1.4}
				& 24.0
				& 12.6
				& 100.0
				& 100.0
				& 16.8 \\
				
				MEP20  & RAND
				& \textbf{0.4}
				& 8.2
				& 5.0
				& 82.8
				& 72.4
				& 1.8 \\
				
				\addlinespace
				
				UNIF20 & EW
				& 24.4
				& 2.4
				& \textbf{0.4}
				& 43.4
				& 24.4
				& 24.8 \\
				
				UNIF20 & VW
				& \textbf{2.4}
				& 7.2
				& 4.6
				& 40.8
				& 30.0
				& 3.4 \\
				
				UNIF20 & SQME
				& 8.8
				& 4.2
				& \textbf{1.0}
				& 55.2
				& 40.6
				& 8.8 \\
				
				UNIF20 & RAND
				& 22.0
				& 2.8
				& \textbf{0.8}
				& 36.8
				& 23.4
				& 21.0 \\
				
				\bottomrule
			\end{tabular}
			
			\vspace{0.4em}
			\begin{minipage}{0.95\textwidth}
				\footnotesize
				\textit{Notes:} Each design contains 500 Buy--and--Hold portfolios.
				Panel A reports the cross-sectional mean of annualized absolute pricing
				errors. Panel B reports the percentage of portfolios with an alpha
				significant at the 5\% level using Newey--West standard errors.
				Boldface denotes the minimum in each row.
			\end{minipage}
		\end{table}
		\FloatBarrier
		
		CAPM and Carhart each produce the smallest mean absolute error in four of the
		eight BH designs. q5 never ranks first. It is close to CAPM in MEP20--EW and
		MEP20--RAND, but deteriorates sharply in MEP20--VW. FF5 and FF6 also leave
		large pricing errors, especially in the MEP20 designs.
		
		Relative to the unfiltered NYSE-only results, q5 improves in four designs and
		worsens in four. The effect of removing the smallest stocks is therefore itself
		conditional on sampling and initial weighting.
		
		\subsection{Constant--Weight}
		
		Table~\ref{tab:app-nyse20-cw} reports the corresponding results when target
		weights are restored daily.
		
		\FloatBarrier
		\begin{table}[H]
			\centering
			\onehalfspacing
			\caption{NYSE Market-Capitalization Breakpoint within the NYSE-Only Universe:
				Constant--Weight}
			\label{tab:app-nyse20-cw}
			\small
			\setlength{\tabcolsep}{5.0pt}
			
			\begin{tabular}{llrrrrrr}
				\toprule
				Sampling & Initial weight
				& CAPM & FF3 & Carhart & FF5 & FF6 & q5 \\
				\midrule
				
				\multicolumn{8}{l}{
					\textit{Panel A: Mean Absolute Pricing Error (bp)}
				} \\
				
				MEP20  & EW
				& 65.7
				& \textbf{37.5}
				& 60.0
				& 110.2
				& 46.7
				& 132.9 \\
				
				MEP20  & VW
				& 64.7
				& 71.0
				& \textbf{23.5}
				& 211.5
				& 146.5
				& 53.0 \\
				
				MEP20  & SQME
				& \textbf{26.5}
				& 48.8
				& 30.9
				& 160.2
				& 92.5
				& 42.1 \\
				
				MEP20  & RAND
				& 68.0
				& \textbf{45.0}
				& 62.5
				& 111.1
				& 51.2
				& 132.0 \\
				
				\addlinespace
				
				UNIF20 & EW
				& 225.7
				& \textbf{46.8}
				& 110.8
				& 68.3
				& 48.5
				& 262.1 \\
				
				UNIF20 & VW
				& 84.7
				& \textbf{74.9}
				& 77.7
				& 132.3
				& 84.4
				& 125.2 \\
				
				UNIF20 & SQME
				& 141.0
				& 45.7
				& 84.6
				& 87.4
				& \textbf{44.5}
				& 197.5 \\
				
				UNIF20 & RAND
				& 223.7
				& 54.7
				& 109.6
				& 75.3
				& \textbf{54.5}
				& 259.2 \\
				
				\midrule
				
				\multicolumn{8}{l}{
					\textit{Panel B: Share of Significant Alphas (\%)}
				} \\
				
				MEP20  & EW
				& 5.0
				& \textbf{1.0}
				& 7.0
				& 46.6
				& 3.8
				& 58.2 \\
				
				MEP20  & VW
				& \textbf{0.0}
				& 5.8
				& \textbf{0.0}
				& 99.8
				& 81.8
				& 0.4 \\
				
				MEP20  & SQME
				& \textbf{0.0}
				& 4.0
				& 0.4
				& 95.2
				& 41.6
				& 1.2 \\
				
				MEP20  & RAND
				& 5.8
				& \textbf{1.4}
				& 7.2
				& 38.8
				& 5.6
				& 46.4 \\
				
				\addlinespace
				
				UNIF20 & EW
				& 62.2
				& \textbf{0.6}
				& 17.8
				& 6.2
				& 1.2
				& 94.8 \\
				
				UNIF20 & VW
				& 4.6
				& \textbf{3.6}
				& 4.4
				& 19.6
				& 5.0
				& 15.4 \\
				
				UNIF20 & SQME
				& 30.2
				& \textbf{0.6}
				& 10.4
				& 13.2
				& 1.2
				& 72.2 \\
				
				UNIF20 & RAND
				& 53.0
				& \textbf{0.6}
				& 15.6
				& 6.2
				& 1.6
				& 87.6 \\
				
				\bottomrule
			\end{tabular}
			
			\vspace{0.4em}
			\begin{minipage}{0.95\textwidth}
				\footnotesize
				\textit{Notes:} Each design contains 500 Constant--Weight portfolios.
				Panel A reports the cross-sectional mean of annualized absolute pricing
				errors. Panel B reports the percentage of portfolios with an alpha
				significant at the 5\% level using Newey--West standard errors.
				Boldface denotes the minimum in each row.
			\end{minipage}
		\end{table}
		\FloatBarrier
		
		The minimum CW errors are distributed across FF3, FF6, Carhart, and CAPM,
		although FF3 has the lowest significant-alpha share in six of the eight designs.
		q5 again never ranks first. It performs relatively well in MEP20--VW and
		MEP20--SQME, but leaves large positive errors and high rejection rates in
		several UNIF20 designs.
		
		Relative to the unfiltered NYSE-only CW results, q5 improves in seven of eight
		designs. The remaining large UNIF20 errors indicate that the smallest stocks
		amplify, but do not fully explain, q5's CW pricing errors.
		
		\subsection{Holding Rules and Dependence-Robust Inference}
		
		The BH--CW comparison is model-specific. FF3, FF5, and FF6 have lower mean
		absolute errors under CW in every design. By contrast, CAPM, Carhart, and q5
		perform better under BH in six of eight designs; q5's exceptions are
		MEP20--VW and MEP20--SQME.
		
		Block-bootstrap inference yields the same contrast. Under BH, the mean alpha of
		FF5 and FF6 is rejected in every design, whereas q5 is rejected in only two.
		Under CW, q5 is rejected in six designs and FF5 in five, while FF3 and Carhart
		are never rejected.
		
		\subsection{Summary}
		
		Excluding the bottom 20\% of NYSE stocks does not remove construction
		dependence. Under BH, CAPM and Carhart divide the cell-level minima, while FF5
		and FF6 retain large pricing errors. Under CW, FF3 generally has the lowest
		significant-alpha share, but no single model minimizes absolute errors across all
		designs.
		
		q5 performs reasonably in some MEP20 cells but leaves large positive errors in
		several UNIF20 and EW or RAND designs. The NYSE-only findings therefore cannot
		be attributed mechanically to microcaps. The breakpoint filter reduces some
		errors, but the dependence on sampling, initial weighting, and BH versus CW
		remains.
		
	\end{appendices}
	

\begin{thebibliography}{99}
		
		\bibitem[Gibbons et al.(1989)]{GRS89}
		Gibbons, M. R., Ross, S. A., \& Shanken, J. (1989).
		A test of the efficiency of a given portfolio.
		\textit{Econometrica}, 57(5), 1121--1152.
		\url{https://www.jstor.org/stable/1913625}
		
		\bibitem[Lo and MacKinlay(1990)]{LM90}
		Lo, A. W., \& MacKinlay, A. C. (1990).
		Data-snooping biases in tests of financial asset pricing models.
		\textit{The Review of Financial Studies}, 3(3), 431--467.
		\url{https://doi.org/10.1093/rfs/3.3.431}
		
		\bibitem[Hansen and Jagannathan(1997)]{HJ97}
		Hansen, L. P., \& Jagannathan, R. (1997).
		Assessing specification errors in stochastic discount factor models.
		\textit{The Journal of Finance}, 52(2), 557--590.
		\url{https://doi.org/10.1111/j.1540-6261.1997.tb04813.x}
		
		\bibitem[Cochrane(2005)]{Cochrane05}
		Cochrane, J. H. (2005).
		\textit{Asset Pricing} (Revised ed.).
		Princeton University Press.
		\url{https://www.johnhcochrane.com/asset-pricing}
		
		\bibitem[Lewellen et al.(2010)]{LNS10}
		Lewellen, J., Nagel, S., \& Shanken, J. (2010).
		A skeptical appraisal of asset-pricing tests.
		\textit{Journal of Financial Economics}, 96(2), 175--194.
		\url{https://doi.org/10.1016/j.jfineco.2009.09.001}
		
		\bibitem[Barillas and Shanken(2017)]{BS17}
		Barillas, F., \& Shanken, J. (2017).
		Which alpha?
		\textit{The Review of Financial Studies}, 30(4), 1316--1338.
		\url{https://doi.org/10.1093/rfs/hhw101}
		
		\bibitem[Barillas and Shanken(2018)]{BS18}
		Barillas, F., \& Shanken, J. (2018).
		Comparing asset pricing models.
		\textit{The Journal of Finance}, 73(2), 715--754.
		\url{https://doi.org/10.1111/jofi.12607}
		
		\bibitem[Kozak et al.(2018)]{KNS18}
		Kozak, S., Nagel, S., \& Santosh, S. (2018).
		Interpreting factor models.
		\textit{The Journal of Finance}, 73(3), 1183--1223.
		\url{https://doi.org/10.1111/jofi.12612}
		
		\bibitem[Giglio et al.(2025)]{GXZ25}
		Giglio, S., Xiu, D., \& Zhang, D. (2025).
		Test assets and weak factors.
		\textit{The Journal of Finance}, 80(1), 259--319.
		\url{https://doi.org/10.1111/jofi.13415}
		
		
		\bibitem[Jegadeesh and Titman(1993)]{JT93}
		Jegadeesh, N., \& Titman, S. (1993).
		Returns to buying winners and selling losers: Implications for stock market efficiency.
		\textit{The Journal of Finance}, 48(1), 65--91.
		\url{https://doi.org/10.1111/j.1540-6261.1993.tb04702.x}
		
		\bibitem[Carhart(1997)]{Carhart97}
		Carhart, M. M. (1997).
		On persistence in mutual fund performance.
		\textit{The Journal of Finance}, 52(1), 57--82.
		\url{https://doi.org/10.1111/j.1540-6261.1997.tb03808.x}
		
		\bibitem[Fama and French(1992)]{FF92}
		Fama, E. F., \& French, K. R. (1992).
		The cross-section of expected stock returns.
		\textit{The Journal of Finance}, 47(2), 427--465.
		\url{https://doi.org/10.1111/j.1540-6261.1992.tb04398.x}
		
		\bibitem[Fama and French(1993)]{FF93}
		Fama, E. F., \& French, K. R. (1993).
		Common risk factors in the returns on stocks and bonds.
		\textit{Journal of Financial Economics}, 33(1), 3--56.
		\url{https://doi.org/10.1016/0304-405X(93)90023-5}
		
		\bibitem[Fama and French(2015)]{FF15}
		Fama, E. F., \& French, K. R. (2015).
		A five-factor asset pricing model.
		\textit{Journal of Financial Economics}, 116(1), 1--22.
		\url{https://doi.org/10.1016/j.jfineco.2014.10.010}
		
		\bibitem[Fama and French(2018)]{FF18}
		Fama, E. F., \& French, K. R. (2018).
		Choosing factors.
		\textit{Journal of Financial Economics}, 128(2), 234--252.
		\url{https://doi.org/10.1016/j.jfineco.2018.02.012}
		
		
		\bibitem[Hou et al.(2015)]{HXZ15}
		Hou, K., Xue, C., \& Zhang, L. (2015).
		Digesting anomalies: An investment approach.
		\textit{The Review of Financial Studies}, 28(3), 650--705.
		\url{https://doi.org/10.1093/rfs/hhu068}
		
		\bibitem[Hou et al.(2019)]{HMXZ19}
		Hou, K., Mo, H., Xue, C., \& Zhang, L. (2019).
		Which factors?
		\textit{Review of Finance}, 23(1), 1--35.
		\url{https://doi.org/10.1093/rof/rfy032}
		
		\bibitem[Hou et al.(2020)]{HXZ20}
		Hou, K., Xue, C., \& Zhang, L. (2020).
		Replicating anomalies.
		\textit{The Review of Financial Studies}, 33(5), 2019--2133.
		\url{https://doi.org/10.1093/rfs/hhy131}
		
		\bibitem[Hou et al.(2021)]{HMXZ21}
		Hou, K., Mo, H., Xue, C., \& Zhang, L. (2021).
		An augmented q-factor model with expected growth.
		\textit{Review of Finance}, 25(1), 1--41.
		\url{https://doi.org/10.1093/rof/rfaa004}
		
		\bibitem[Hou et al.(2024)]{HMXZ24}
		Hou, K., Mo, H., Xue, C., \& Zhang, L. (2024).
		The economics of security analysis.
		\textit{Management Science}, 70(1), 164--186.
		\url{https://doi.org/10.1287/mnsc.2022.4640}
		
		
		\bibitem[French(2026)]{FrenchDataLibrary}
		French, K. R. (2026).
		\textit{Kenneth R. French Data Library} [Data set].
		Accessed May 5, 2026.
		\url{https://mba.tuck.dartmouth.edu/pages/faculty/ken.french/data_library.html}
		
		\bibitem[Global-q.org(2026)]{globalqFactors}
		Global-q.org. (2026).
		\textit{Factors and testing portfolios} [Data set].
		Accessed May 5, 2026.
		\url{https://global-q.org/factors.html}
		
	\end{thebibliography}
\end{document}